\documentclass[prb,amsmath,superscriptaddress,twocolumn]{revtex4}

\usepackage{epsfig}
\usepackage{amssymb}
\usepackage{amsmath}
\usepackage{amsfonts}
\usepackage{bbold}
\usepackage{bm}
\usepackage{graphicx}
\usepackage{soul}
\usepackage{xcolor}
\usepackage{hyperref}

\newcommand{\nn}{\nonumber}
\newcommand{\ud}{{\textrm{d}}}

\begin{document}

\title{Quasi-Many-Body Localization of Interacting Fermions with Long-Range Couplings}

\author{S. J. Thomson}
\email{steven.thomson@polytechnique.edu}
\affiliation{Centre de Physique Th\'{e}orique, CNRS, Institut Polytechnique de Paris, Route de Saclay, F-91128 Palaiseau, France}
\affiliation{Institut de Physique Th\'eorique, Universit\'e Paris-Saclay, CNRS, CEA, F-91191 Gif-sur-Yvette, France}
\author{M. Schir\'o}
\email{marco.schiro@ipht.fr}
\thanks{On Leave from: Institut de Physique Th\'{e}orique, Universit\'{e} Paris Saclay, CNRS, CEA, F-91191 Gif-sur-Yvette, France}
\affiliation{JEIP, USR 3573 CNRS, Coll\`ege de France, PSL Research University, 11 Place Marcelin Berthelot, 75321 Paris Cedex 05, France}

\date{\today}

\begin{abstract}
A number of experimental platforms for quantum simulations of disordered quantum matter, from dipolar systems to trapped ions, involve degrees of freedom which are coupled by power-law decaying hoppings or interactions, yet the interplay of disorder and interactions in these systems is far less understood than in their short-ranged counterpart. Here we consider a prototype model of interacting fermions with disordered long-ranged hoppings and interactions, and use the flow equation approach to map out its dynamical phase diagram as a function of hopping and interaction exponents. We demonstrate that the flow equation technique is ideally suited to problems involving long-range couplings due to its ability to accurately simulate very large system sizes. We show that, at large on-site disorder and for short-range interactions, a transition from a delocalized phase to a quasi many-body localized (MBL) phase exists as the hopping range is decreased. 
This quasi-MBL phase is characterized by intriguing properties such as a set of emergent conserved quantities which decay algebraically with distance. Surprisingly we find that a crossover between delocalized and quasi-MBL phases survives even in the presence of long-range interactions.
\end{abstract}

\maketitle
\section{Introduction}
Recent years have seen tremendous progress in our understanding of how isolated quantum many-body systems approach thermal equilibrium or fail to do so, sparking great interest in the possibility of engineering exotic non-ergodic phases of quantum matter~\cite{RigolETH08,DAlessioEtAl2016,Nandkishore+15,AbaninEtAlRMP19}. The interest around this question has substantially broadened across disciplines, evolving from a purely speculative issue in the foundation of quantum statistical mechanics~\cite{DeutschPRA91} to a central topic of modern research, from condensed matter~\cite{AltmanNaturePhysics2018} to high-energy physics~\cite{QiXLNatPhys2018,DymarskyPavlenkoPRL19}, with direct implications for the robustness of future quantum technologies. In particular, quantum ergodicity breaking may pave the way towards novel platforms to store and protect quantum information from intrinsic decoherence~\cite{Huse+13,Bahri+15}, a development with clear technological signifiance. 

Among possible scenarios for ergodicity breaking, special attention has been devoted in the recent past to the role of quenched disorder and interactions, leading to Many-Body Anderson Localization (MBL)~\cite{Anderson58,Gornyi+05,Basko+06,Pal+10,Alet+18}.  Experimental advances in quantum simulators have allowed unprecedented control over disordered many-body systems and reported evidence of MBL behavior in a number of platforms, ranging from one and two dimensional arrays of ultracold atoms~\cite{BlochMBL2015,Kondov+15,Choi+16,RispoliEtAlNature19,LukinScience2019} to ion traps with programmable random disorder~\cite{Smith+16,Zhang+17} and dipolar systems made by nuclear spins~\cite{Alvarez846,KucskoEtAlPRL18}. Interestingly, most of the relevant platforms for quantum simulations of disordered many body systems involve degrees of freedom which are coupled by long-range hopping processes or interactions, typically decaying as a power-law of the distance. While the interplay of disorder and interaction leading to MBL is by now rather well understood for one-dimensional models with short-range interactions, where a set of mutually commuting, exponentially localized integrals of motion (LIOMs, or $l$-bits) can be identified~\cite{SerbynPapicAbaninPRL13_2,HuseNandkishoreOganesyanPRB14,Imbrie16a,Imbrie+16b} its fate in the presence of long-range couplings 
is far less settled.  From one side, perturbative arguments suggest an instability of the MBL phase in quantum spin chains with interactions of random sign~\cite{Burin06,Yao+14,Burin15,Gutman+16} decaying with an exponent $\beta<2d$ (with $d$ the spatial dimension of the system), while avalanche arguments~\cite{DeRoeckHuveneersPRB17} would rule out a genuine MBL behavior for interactions decaying slower than exponential, as do numerical simulations of spin transport close to the MBL transition \cite{Kloss+19}. On the other hand, experiments continue to find evidence of localization in this regime \cite{Smith+16,Zhang+17,Alvarez846,KucskoEtAlPRL18}, and several scenarios have recently emerged which are consistent with localized behavior even for slowly decaying power laws~\cite{Nandkishore+17,Santos+16,Roy+19,DengEtAlArxiv19,DengEtAlPRL18,NosovEtAlPRB19}. Exact diagonalization, which played a crucial role in understanding conventional short-ranged MBL, is limited to small sizes and suffers from strong finite size effects in long-range models, making the theoretical descriptions of disordered interacting quantum systems with power-law couplings a major open challenge, whose solution is particularly pressing given the experimental evidence of quasi-MBL in a number of quantum simulators at the interface between solid state and atomic physics.

In this work we address this problem for a model of interacting fermions where both hopping and interaction are disordered and power-law decaying, with different exponents. Using a significantly improved and extended variant of the truncated flow equation approach, already proven to be able to describe both the short-ranged MBL phase in both one and two dimensions~\cite{Thomson+18} and the well understood delocalization of non-interacting fermions with power-law hopping~\cite{Thomson+20}, we map out the static and dynamical properties of the system as a function of the hopping and interaction exponents. We find that for rapidly decaying power laws the system at large on-site disorder is in a quasi-MBL phase \footnote{We use the term ``quasi-MBL'' to emphasize that this phase might be metastable and that on longer times, or at larger system sizes avalanche instability would lead to a complete delocalization - see Ref. \onlinecite{DeRoeckHuveneersPRB17} and the Discussion at the end of our manuscript for more details.} characterized by algebraically decaying $l$-bit interactions~\cite{DeTomasiPRB19,DengEtAlArxiv19} that we explicitly construct. Remarkably, the flow equation technique is able to capture the delocalization of this quasi-MBL phase upon decreasing the hopping exponent, a non-trivial result that confirms the reliability of this approach. Surprisingly we find that the quasi-MBL phase survives upon increasing the range of the interactions, though with a significantly broadened crossover to the ergodic regime. We speculate that this phase may be unstable in the thermodynamic limit, and discuss possible connections with other works.

The paper is organised as follows. In Section~\ref{sec.model} we first describe the model we propose, and discuss how it links to other models studied in the literature. In Section~\ref{sec.method}, we discuss in detail the flow equation method which we use, and in Section~\ref{sec:benchmarks} we provide detailed benchmarks for both static and dynamic quantities to demonstrate the high accuracy that can be achieved by this technique. In Section~\ref{sec.lbits} we present results for the local integrals of motion computed using this method, as well as the coupling constants of the fixed-point Hamiltonian, and show that they behave markedly differently. In Section~\ref{sec.dyn}, we go on to compute the non-equilibrium dynamics using flow equations, presenting results for the imbalance and a complete phase diagram. We end with a discussion in Section~\ref{sec.dis} and conclude with an outlook towards the future in Section~\ref{sec.con}, and finally include a series of technical Appendices which include additional details and comparisons with other disorder distributions.

\section{The model}
\label{sec.model}
Theoretical investigations of localization in long-range systems date back to Anderson's original work \cite{Anderson58}. One well-understood example is the non-interacting random hopping problem, where the hopping terms decay as a power-law with exponent $\alpha$, also known as Power-Law Random Banded Matrix (PRBM) model. In this case, localization is destroyed for $\alpha<d$ (where $d$ is the spatial dimension) and the system is critical at $\alpha=d$ \cite{Yeung+87,Levitov90,Mirlin+96,Levitov99,Varga+00,Mirlin+00,Evers+00,Kravtsov+05,Evers+08}. Here, we wish to study an interacting variant of the PRBM model, incorporating random long-range interactions in addition to the random long-range hopping terms.  We therefore consider a Hamiltonian describing a one-dimensional chain of interacting fermions given by: 
\begin{align}\label{eqn:model}
\mathcal{H} =  \sum_i h_i n_i + \frac12 \sum_{ij} V_{ij} n_i n_j + \sum_{ij} J_{ij} c^{\dagger}_i c_{j}
\end{align}
where the on-site disorder is drawn from a box distribution $h_i \in [0,W]$. The couplings $J_{ij}=J_{ji}$ and $V_{ij}=V_{ji}$ are also random and drawn from Gaussian distributions with zero mean and standard deviations which decay with distance as $\sigma_{J}= J_0/|i-j|^{\alpha}$ and $\sigma_{V}= V_0/|i-j|^{\beta}$ respectively. Unless otherwise specified, we fix $J_0 = 0.5$, $V_0 = 0.1$ and  $W=5$, such that the model with short-ranged hopping and interactions (respectively $\alpha=\beta=\infty$) would be in the MBL phase, and vary the power-law exponents $\alpha$ and $\beta$ only.

To our knowledge, this model has not been studied in the literature before. In Ref.~\onlinecite{Khatami+12}, a related model of interacting fermions with random power-law hopping was studied numerically, but the role of on-site disorder and random, power-law interactions was not considered. Interestingly, in the $\alpha,\beta\rightarrow0$ limit, Eq.~(1) reduces to a model of fermions with all-to-all random couplings, reminiscent of the maximally chaotic Sachdev-Ye-Kitaev model~\cite{SachdevYePRL93} with the addition of a random, on-site disorder. In the literature,  several studies have focused on quantum spin models with power-law decaying exchange couplings of random signs, which however are not equivalent to fermionic models due to the long-range nature of the couplings.  For these models estimates based on the locator expansion and its breakdown suggest an instability of the (many-body) localised phase for slowly decaying transverse exchange with exponent $\beta <2d$~\cite{Burin06,Yao+14,Burin15}, independently of the longitudinal exponent $\alpha$ which controls the degrees of freedom involved in resonance formation~\cite{Yao+14,Gutman+16}. The robustness and generality of those perturbative arguments however has not been fully discussed.  In particular, convergence of the locator expansion provides at most a sufficient condition for localization but does not usually guarantee delocalization. Different scenarios have emerged recently which are consistent with localised behavior even in presence of slowly decaying power-law interactions, for which the locator expansion does not converge. Examples include order-enabled localization~\cite{Nandkishore+17} cooperative shielding~\cite{Santos+16,Roy+19,DengEtAlArxiv19}, correlation-induced localization in single particle problems~\cite{DengEtAlPRL18,NosovEtAlPRB19} or the existence of a critical disorder for localization at finite size~ \cite{Tikhonov+18}, vanishing in the thermodynamic limit.

\section{Method}
\label{sec.method}
Systems with long-range couplings are typically extremely challenging to study numerically, as they require very large system sizes in order to avoid finite-size effects as the interaction range is increased. With the addition of disorder in the long-range couplings, the model in Eq.~(\ref{eqn:model}) falls into a class of systems which cannot be efficiently simulated using Matrix Product State methods, where long-range couplings are typically represented as a sum of decaying exponentials, which is not straightforward for \emph{disordered} long-range couplings. As a consequence a vast majority of numerical results rely on  exact diagonalisation (ED), which in a non-sparse model with long-range couplings is limited to small system sizes where finite-size effects will be significant. 

To address this challenging problem here we make use of the flow equation approach \cite{Wegner94,Kehrein07,Moeckel+08,Hackl+08,Hackl+09,Eckstein+09,Monthus16,Quito+16,Pekker+17,Savitz+19,Yu+19,Kelly+20} which we have recently used to study MBL in the short ranged case~\cite{Thomson+18} as well as the non-interacting PRBM model~\cite{Thomson+20} and in a periodically driven Floquet system with weak interactions~\cite{thomson2020flow}.

The main idea is to diagonalize the Hamiltonian through a series of {\it infinitesimal} unitary transforms parametrised by a fictitious `flow time' $l$ which runs from $l=0$ (initial basis) to $l \to \infty$ (diagonal basis). The Hamiltonian flow reads
\begin{align}\label{eq.flow}
\frac{\ud \mathcal{H}}{\ud l} = [\eta(l),\mathcal{H}(l)].
\end{align}
where $\eta(l)$ is the generator of the flow and the initial condition at $l=0$ is given by the Hamiltonian in Eq.~(\ref{eqn:model}).
In the following, we shall use Wegner's choice of generator \cite{Wegner94} $\eta(l)=[\mathcal{H}_0(l),V(l)]$, where $\mathcal{H}_0$ contains the terms which are diagonal in a given basis, while $V$ contains the off-diagonal terms. This choice of generator, although not unique~\cite{Monthus16,Savitz+17,Thomson+20}, guarantees \cite{Wegner94,Kehrein07} that the off-diagonal terms vanish in the $l \to \infty$ limit. While for quadratic problems the flow equation approach is exact, in the presence of interactions the flow generates higher-order couplings not present in the original microscopic model.
To deal with these, we use a truncation scheme, originally introduced in Ref~\onlinecite{Thomson+18}, that we briefly discuss  below for the present case.

\subsection{Generator of the Flow and Truncation}

 We make an ansatz for the form of the running Hamiltonian $\mathcal{H}(l) = \mathcal{H}_0(l) + V(l)$, with 
\begin{align}\label{eqn:ansatz}
\mathcal{H}_0(l)& =  \sum_i h_i(l) :c^{\dagger}_i c_i: + \frac12 \sum_{ij} \Delta_{ij}(l) :c^{\dagger}_{i} c_{i} c^{\dagger}_{j} c_{j}: \\
V(l) &= \sum_{ij} J_{ij}(l) :c^{\dagger}_i c_{j}:,
\end{align}
where the $:\mathcal{O}:$ notation signifies normal-ordering. We adopt normal ordering using the $:\hat{O}:$ notation in order to i) ensure a consistent ordering of operators when computing commutation relations, and ii) efficiently resum contributions from higher-order terms to turn the flow equation method into a powerful non-perturbative scheme - see Refs. \onlinecite{Wegner06,Kehrein07} and Appendix A for details.  Given the ansatz above the Wegner generator reads
\begin{align}
\eta &= \sum_{ij} \mathcal{F}_{ij} :c^{\dagger}_i c_j: + \sum_{ijk} \zeta^{k}_{ij} :c^{\dagger}_k c_k c^{\dagger}_i c_j:
\end{align}
with $\mathcal{F}_{ij} \equiv J_{ij} \left[ (h_i - h_j) - \Delta_{ij} (\langle n_i \rangle - \langle n_j \rangle)  \right]$ and $\zeta^k_{ij} \equiv J_{ij} (\Delta_{ik} - \Delta_{jk})$, where the scale-dependence of the coefficients has been suppressed for clarity.

The flow of the Hamiltonian is given by Eq.~\ref{eq.flow}. using the expressions above, it can be clearly seen that the commutation relation between the interaction term of the Hamiltonian and the interacting part of the generator will lead to the generation of new higher-order terms in the Hamiltonian during the flow.
In practice, the successive generation of these higher-order terms quickly renders the calculation analytically intractable, however for weak interactions the newly-generated terms have only an extremely small spectral weight. Specifically, the lowest-order commutator responsible for generating new higher-order terms has the following form:
\begin{align}
\sum_{ijk} \sum_{lm} J_{ij}(\Delta_{ik} - \Delta_{jk}) \Delta_{lm} [:c^{\dagger}_k c_k c^{\dagger}_i c_j:, :n_l n_m:].
\end{align}
The result of this term will be at maximum of order $J_0 V_0^2$, and as $V_0 \ll 1$, the generation of high order terms is heavily suppressed and this term may be considered negligible. We therefore discard all newly generated terms and restrict ourselves to the variational manifold.  Thus, we can conclude to a high degree of certainty that this truncation is accurate for the weak interactions considered here. Crucially, we can monitor the accuracy of our truncation scheme, as we discuss further in Section~\ref{sec:benchmarks}.

\subsection{Flow Equations}
The flow of the Hamiltonian coefficients can be read off from $\ud \mathcal{H} / \ud l = [\eta(l), \mathcal{H}(l)]$, following a lengthy calculation. Explicit expressions for the flow equations are as follows:
\begin{align}
\frac{\ud h_i (l)}{\ud l} &= 2 \sum_j J_{ij}^{2}(h_i - h_j) - 4 \sum_j J_{ij}^2 \Delta_{ij} (\langle n_i \rangle - \langle n_j \rangle) \nn\\
& + \sum_{jk} J_{jk}^2 (\Delta_{ik} - \Delta_{ij}) (\langle n_k \rangle - \langle n_j \rangle) \\
\frac{\ud J_{ij}(l)}{\ud l} &= -J_{ij}(h_i - h_j)^2 - \sum_{k} J_{ik}J_{kj}(2h_k - h_i - h_j) \nn\\
&+ 2J_{ij} \Delta_{ij}(h_i - h_j)(\langle n_i \rangle - \langle n_j \rangle) \nn \\
& \quad -J_{ij} \Delta_{ij}^2 \left( \langle n_i \rangle + \langle n_j \rangle - 2 \langle n_i \rangle \langle n_j \rangle \right) \nn\\
& - \frac12 \sum_k J_{ij} (\Delta_{ik}-\Delta_{jk})^2 \langle n_k \rangle (1 - \langle n_k \rangle) \label{eq.flowJ} \nn \\
& + \sum_k J_{ik} J_{kj} \left[ (\Delta_{ij} - 2 \Delta_{jk})(\langle n_j \rangle - \langle n_k \rangle) \right. \nn \\
& \left. + (\Delta_{ij} - 2 \Delta_{ik})(\langle n_i \rangle - \langle n_k \rangle) \right] \\
\frac{\ud \Delta_{ij}(l)}{\ud l} &= 2 \sum_{k \neq i,j} \left[ J_{ik}^2 (\Delta_{ij} - \Delta_{kj}) + J_{jk}^2 (\Delta_{ij} - \Delta_{ik}) \right] \label{eq.flowD}
\end{align}
In the $l \to \infty$ limit, the off-diagonal terms $J_{ij}$ vanish and we obtain a diagonal Hamiltonian given by
\begin{align}\label{eq.Htilde}
\tilde{\mathcal{H}} = \sum_i \tilde{h}_i n_i + \frac12\sum_{ij} \tilde{\Delta}_{ij} n_i n_j
\end{align}
In all of the following, the tilde notation indicates quantities in the $l \to \infty$ diagonal basis.  In practice, we numerically integrate these equations until the off-diagonal elements have decayed to the required accuracy, typically using $l_{max} \approx 10^3$ and discarding couplings which have reached zero below some cutoff (typically $10^{-6}$ or less). In cases where the flow is slow to converge, e.g. the weak-disorder limit, Eq. \ref{eq.flowD} can exhibit spurious divergences which must be handled carefully in order to obtain physically reasonable results. The consequences of this divergence is that the normal-ordering corrections in Eq. \ref{eq.flowJ} can contribute an unphysically large negative contribution to the flow of the off-diagonal elements, effectively sending them to zero exponentially quickly as the system of equations attempts to stop the divergence, resulting in a deviation from unitarity. In order to maintain an accurate flow in this regime, one can monitor the flow equations at each flow time step and if a divergence occurs, subtract both the divergent term in Eq. \ref{eq.flowD} and its counter term in Eq. \ref{eq.flowJ}. This has the effect of `freezing' the divergent terms while still allowing the other terms to continue flowing. We note, however, that this is typically not a problem in the strong-disorder regime we consider here.

\subsection{Non-Equilibrium Dynamics}

In addition to obtaining the fixed point Hamiltonian and its approximated spectrum, restricted to the ansatz in Eq.(\ref{eqn:ansatz}), we can also compute the real-time dynamics of an operator by transforming it into the basis which diagonalises the Hamiltonian, time-evolving with respect to the diagonal Hamiltonian, and then flowing the operator back into the physical basis. We discuss this in detail for the number operator $n_i(t)$ whose dynamics will be presented in Section~\ref{sec.results}.

To parameterise the flow of this operator, we make the following ansatz for the {\it running} number operator at time $t=0$
\begin{align}
n_i(l,t=0) &= \sum_{j} A^{(i)}_j(l) n_j + \sum_{jk} B^{(i)}_{jk}(l) c^{\dagger}_j c_k \label{eq.nansatz}
\end{align}
with initial conditions $A^{(i)}_j(l=0) = \delta_{ij}$ and $B^{(i)}_{jk}(l=0) = 0 \phantom{.} \forall \phantom{.} j,k$. The flow equations for this operator can be obtained by computing $dn_i(l)/dl = [\eta(l), n_i(l)]$ and are given by:
\begin{align}
\frac{\ud A^i_j}{\ud l} &= -2 \sum_k J_{jk} (h_k - h_j) B_{kj}, \label{eq.flow_A} \\
\frac{\ud B_{jk}}{\ud l} &= -J_{jk} (h_k-h_j)(A^i_k-A^i_j) \nn\\
& \quad -\sum_{n} \left[ J_{nj} (h_n - h_j) B_{nk} + J_{nk} (h_n - h_k) B_{nj} \right],  \label{eq.flow_B} 
\end{align}
Note that higher-order terms cannot be consistently included at this order of the truncation scheme, as their flow is constrained by terms not included in the ansatz for the running Hamiltonian. One may attempt to include higher order terms in Eq.~\ref{eq.nansatz} even without the corresponding terms in the Hamiltonian, however in this case we find that they are typically poorly controlled and often divergent. The normal-ordering procedure employed as part of this construction (see Appendix~\ref{sec.normal_ordering}) does, however, allow us to take into account the leading effects of the interactions even at this order.
After transforming $n_i(t=0)$ into the diagonal basis, by solving Eq.~(\ref{eq.flow_A}-\ref{eq.flow_B}) from $l=0$ up to $l=\infty$, we can time-evolve it with respect to the diagonal Hamiltonian~(\ref{eq.Htilde}). As this is still interacting, despite being diagonal, the exact time evolution would require to sum over the exponentially many classical configurations spanned by $n_i=\left\{0,1\right\}$, for every $i$,  which is not practical for large system sizes. Instead, we proceed by writing down the Heisenberg equations of motion and performing a  time-dependent decoupling of the interaction term to get
\begin{align}
\tilde{n}_i(l = \infty, t) =  \sum_{j} A^{(i)}_j(l) n_j + \sum_{jk} B^{(i)}_{jk}(l) \textrm{e}^{i \phi_{jk}(t)} c^{\dagger}_j c_k \\
\phi_{jk}(t) = \int_0^t \ud t' \left[ (\tilde{h}_k - \tilde{h}_j) + \sum_m (\tilde{\Delta}_{km} - \tilde{\Delta}_{jm}) \langle n_m(t') \rangle \right] \label{eq.meanfield}
\end{align}
where the expectation values are calculated self-consistently at each timestep, an approach which represents a significant improvement upon the previous version of this method presented in Ref.~\onlinecite{Thomson+18}. We then use the flow equations (Eqs.~\ref{eq.flow_A} and \ref{eq.flow_B}) to transform the number operator back into the original basis, where it will take the form:
\begin{align}
n_i(l=0,t) &= \sum_{j} A^{(i)}_j(t) n_j + \sum_{jk} B^{(i)}_{jk}(t) c^{\dagger}_j c_k
\end{align}
where the $A^{(i)}_j(t)$ terms picks up an implicit time-dependence during the transform back into the initial basis. At this point, the expectation value of this operator may be computed with respect to the desired initial state. 

\section{Benchmarks}
\label{sec:benchmarks}

In this section we present, for the model defined in Eq.~(\ref{eqn:model}), detailed benchmark results of the flow equation method. Specifically we compare the flow equation results with exact numerics on small system sizes for eigenstates and dynamics. Furthermore we assess the validity of the truncation scheme discussed in Section ~\ref{sec.method} by monitoring the conservation of the so called flow-invariants. The readers interested more in the physics of the problem~(\ref{eqn:model}) and the interplay between MBL and power-law couplings,  can directly jump to Section~\ref{sec.results}.

\subsection{Eigenvalue Comparison with Exact diagonalization}
\begin{figure}[h!]
\begin{center}
\includegraphics[width= \linewidth]{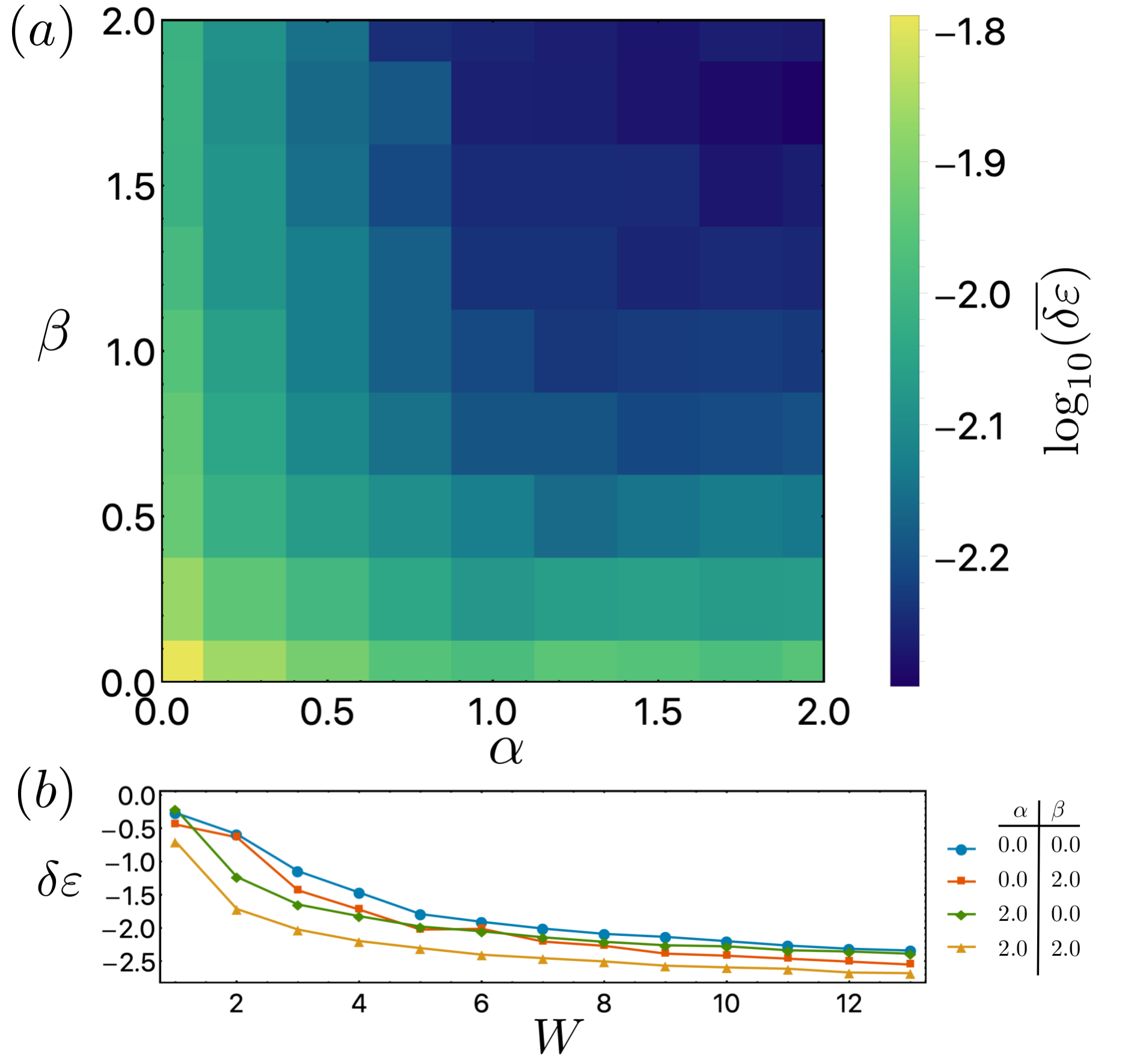}
\caption{The logarithm of the disorder-averaged relative error in the eigenvalues $\overline{ \delta \varepsilon}$ computed with respect to exact diagonalization. (a) The relative error plotted across the same parameter values as the phase diagram in Fig.~\ref{fig.phase} and averaged over $N_s=512$ disorder realizations. The error is largest in the case where all couplings are both long-range, and decreases sharply when either or both exponents have a value greater than zero. Note, however, that the average error remains extremely small across the entire parameter region. 
(b) The disorder-averaged relative error plotted for four fixed values of $(\alpha,\beta)$ against the on-site disorder strength $W$. In the remainder of this work, we fix $W=5$, however here we show how the relative error decreases as the system becomes more strongly disordered.
}
\label{fig.err}
\end{center}
\end{figure}

We first compare the static properties (i.e. the eigenvalues) for a small system of size $L=12$ with Exact diagonalization (ED) results obtained using the QuSpin package \cite{Weinberg+17,Weinberg+19}. We define the averaged relative error as:
\begin{align}
\delta \varepsilon = \frac{1}{N} \sum_i^N \frac{| \varepsilon^{FE}_i - \varepsilon^{ED}_i|}{\varepsilon^{ED}_i}
\end{align}
where $\varepsilon^{FE/ED}$ refer to the many-body eigenvalues obtained using flow equations (FE) and ED methods respectively, and the sum runs over states in the many-body Hilbert space. We can compute this quantity, here restricting ourselves to the half-filled states, for a variety of power-law exponents $\alpha$ and $\beta$ in order to benchmark the accuracy of our results. The results are summarised in Fig. \ref{fig.err}, where we show the average relative error across the parameter range we will consider in this work, here for a system size of $L=12$ and with $N_s = 512$ disorder realisations. We also verified that the error decreases rapidly with increasing disorder strength, as expected, shown in Fig. \ref{fig.err}b). We note that it is almost always possible to reduce the error further by increasing the maximum flow time $l_{max}$, however as the method asymptotically approaches the exact eigenvalues we see diminishing returns by increasing the flow time further, compared with the increased CPU time required to obtain the results.

\begin{figure}[t]
\begin{center}
\includegraphics[width= \linewidth]{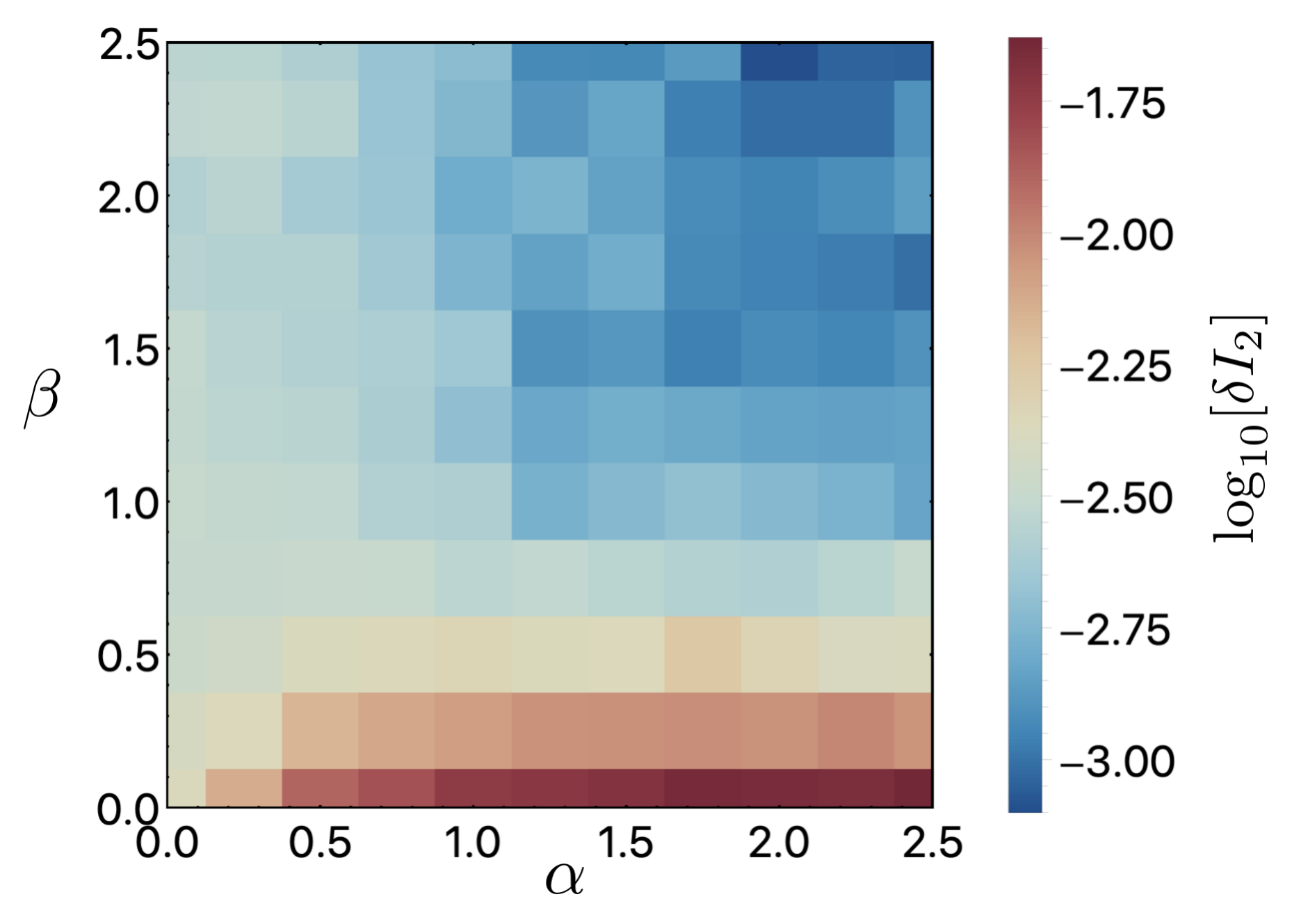}
\caption{Behaviour of the flow invariant across the phase diagram, with $L=64$ and $W=5$. The flow invariant is maximal for $\beta = 0$. Note that the colour scale shows the {\it logarithm} of $\delta I_2$: the deviation of the flow equation transform from perfect unitarity is less than one percent across the majority of the phase diagram. Each of the $11 \times 11$ points in this phase diagram is the result of $50 \leq N_s \leq 128$ disorder realizations, as required for convergence. 
}
\label{fig.invar}
\end{center}
\end{figure}

\begin{figure*}
\begin{center}
\includegraphics[width= \linewidth]{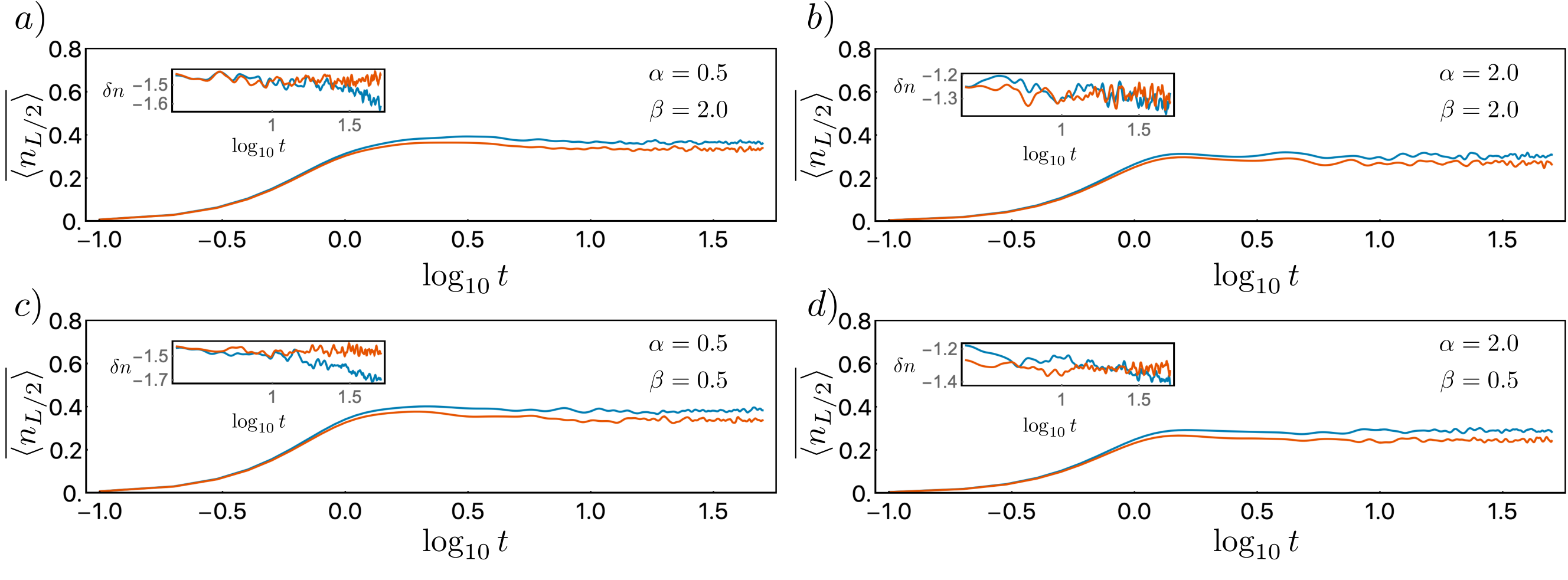}
\caption{Benchmarks of the density dynamics on the central site of a chain of length $L=12$ when quenched from a CDW initial state and averaged over 512 disorder realizations, comparing ED (blue) with FE (orange). a) $\alpha = 0.5, \beta = 2.0$, b) $\alpha = 2.0, \beta=2.0$, c) $\alpha=0.5, \beta=0.5$. d) $\alpha = 2.0, \beta = 0.5$. In all cases, the results are close, but the FE method slightly overestimates the localization. In the more strongly localized regime for $\alpha, \beta \gg 1$, the FE and ED results agree very closely. The insets show the decay of fluctuations around their long-time mean value, with $\delta n = \overline{\sigma^2(\langle n_{L/2} \rangle - \tilde{n} )}$ and $\tilde{n} = \overline{\langle n_{L/2}(t) \rangle}_{t \to \infty}$: note the power-law decay in the ED data is not seen in the FE data, due to the mean-field decoupling employed.}
\label{fig.EDdyn}
\end{center}
\end{figure*}

\subsection{Invariants of the Flow}

As with any other unitary transform, there are a variety of conserved quantities of the flow equation formalism. Specifically, traces of integer powers of the Hamiltonian $I_p = \textrm{Tr} [\mathcal{H}^p]$ are commonly known as `invariants of the flow', and are preserved by an exact implementation of the flow equation formalism. As we have seen, however, in order for the calculation to remain tractable we must make an approximation for the running Hamiltonian of the system. The neglect of any terms not contained within the ansatz Hamiltonian introduces an error: this error may be quantified by computing the invariants of the flow at the start and end of the procedure, and then computing the difference between them. This difference is zero if the unitary transform is exact, and non-zero if the truncation has introduced an error. This allows us to have a self-consistent estimate of the error in the transform which we can compute for any system size, in addition to the relative error measured with respect to ED which we can only compute on small system sizes accessible to exact numerical methods.
Here we focus on the second invariant~\cite{Monthus16} $(p=2)$ and define the truncation error as:
\begin{align}
\delta I_2 = \frac{ | I_2(l=0) - I_2(l = \infty) | } {\frac12 (I_2(l=0) + I_2(l = \infty))}
\end{align}

The main source of error in this scheme is the strength of the interactions, which contribute to the generation of higher-order terms not included in our variational manifold. In the present case, as the truncated higher-order terms scale approximately with integer powers of the interaction strength $V_0 \ll 1$, the neglected terms are typically small and the accuracy very good. However, in the limit of $\beta \to 0$, there are a large number of interaction terms and the neglected terms can begin to become significant. To get an idea of the accuracy of our results, we can compute this quantity across the phase diagram in the $(\alpha,\beta)$ plane: the result is shown in Fig. \ref{fig.invar}. We find that the transform is almost perfectly unitary across the entire phase diagram, with the main deviations away from unitarity occuring close to $\beta = 0$.

\subsection{Comparison with Exact Dynamics}

Finally, in order to verify the accuracy of the time evolution obtained with flow-equations we benchmark it with exact quantum dynamics (ED). For this, we again employed the QuSpin package \cite{Weinberg+17,Weinberg+19}. Sample results for the density dynamics on a single site are shown in Fig. \ref{fig.EDdyn} for a variety of values of $\alpha$ and $\beta$ across the phase diagram. The agreement in all cases is excellent, with flow equations differing only very slightly from the exact results.

Despite this striking agreement of the averaged density dynamics, it is interesting to note that the results from the flow equation method do not capture the decay of fluctuations around their mean values (shown in the insets of Fig. \ref{fig.EDdyn}). The reason for this is due to the mean-field decoupling used in Eq.~\ref{eq.meanfield}, which does not allow for the slow build-up of correlations that leads to the power-law decay of fluctuations (or to the logarithmic growth of entanglement entropy). Similar results are seen in the quantum Fisher information (not shown), a proxy for the entanglement entropy, which does not display the expected slow increase with time due to the nature of the mean-field decoupling used here in computing the dynamics.

\section{Results}
\label{sec.results}

We are now in position to present the main results of this work, concerning the effect of long-range couplings on MBL physics as encoded in the model in Eq.~(\ref{eqn:model}). In the following we focus on the behaviour of this model in the weakly-interacting regime (unless otherwise specified, we fix $J_0 = 0.5$ and $V_0 = 0.1$ and  $W=5$)  with $0 \leq \alpha, \beta \leq 2d$ and study the interplay/competition between power-law hoppings and power-law interactions. We first consider the two effects separately, fixing $\alpha=\infty$ and varying $\beta$ and vice versa, while later we present a complete phase diagram in the $(\alpha,\beta)$ plane.

\begin{figure}[h!]
\begin{center}
\includegraphics[width= \linewidth]{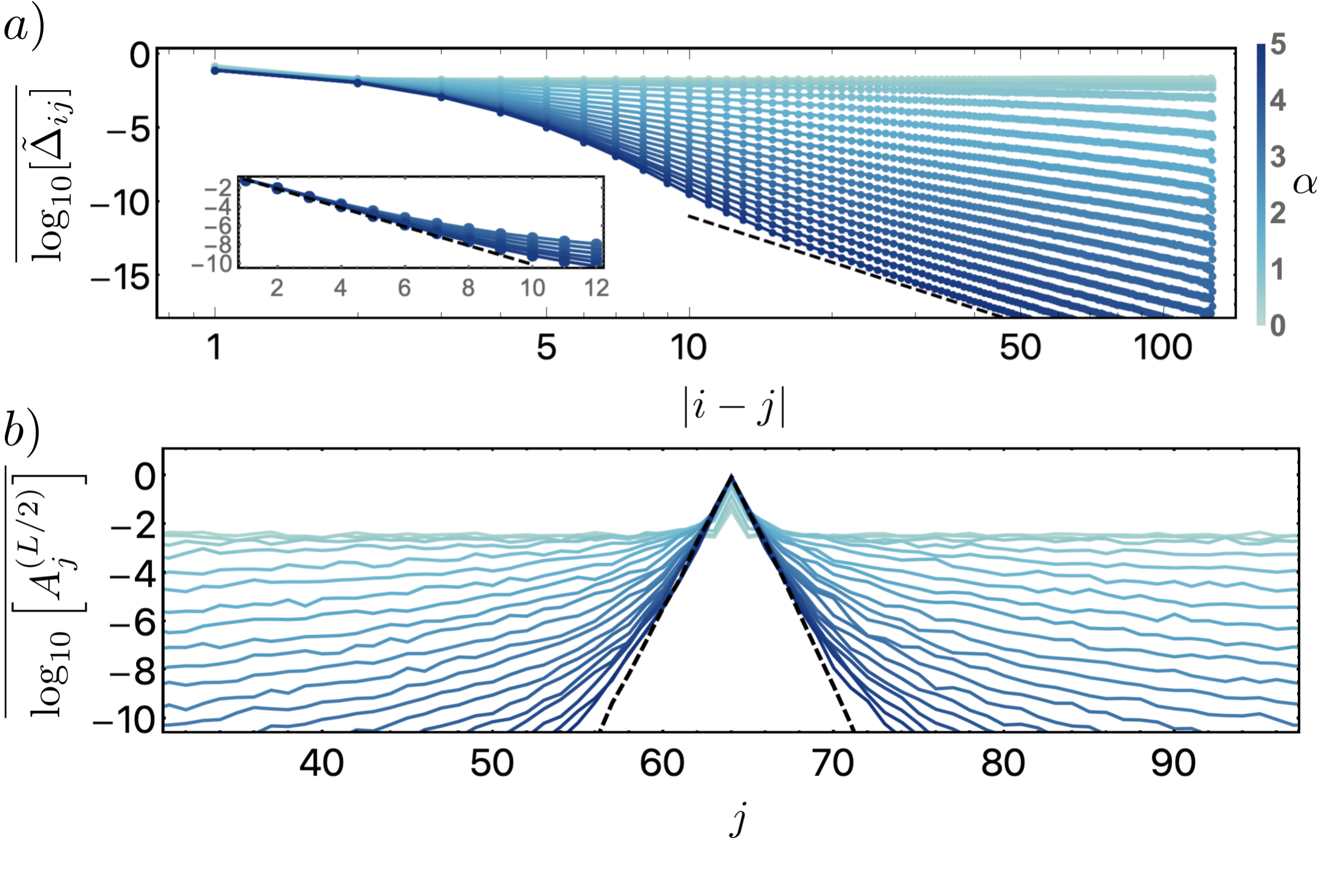}
\caption{$l$-bit interactions (top) and real-space support (bottom) for power-law hopping and nearest-neighbor interactions ($\alpha \in [0.0,5.0]$ (top to bottom) in increments of $0.25$, and $\beta = \infty$). a) The disorder-averaged (median) $\tilde{\Delta}_{ij}$ decay as a power-law at long distances (notice log-log scale, dashed line is a power law guide to eye) and as an exponential at short distances (see inset, semi-log scale, for $\alpha \in [3.5,5]$). b) The $l$-bits exhibit an exponential decay (most visible for large $\alpha$) crossing over to an extended behavior with long power-law tails. The dashed line is the $(\alpha \to \infty, \beta \to \infty)$ short-range limit. Chain size $L=128$, disorder realizations $N_s=256$.}
\label{fig.pwrhop}
\end{center}
\end{figure}

\subsection{Decay of $l$-bit interactions and real-space support} 
\label{sec.lbits}

We start discussing the properties of the fixed point diagonal Hamiltonian~(\ref{eq.Htilde}) obtained by solving the flow-equations. This describes a model of localized bits (or $l$-bits) in presence of random fields $\tilde{h}_i$ and pairwise interactions  $\tilde{\Delta}_{ij}$.  First, we can straightforwardly extract the distance dependence of the coefficients $\tilde{\Delta}_{ij}$, as our procedure automatically generates the Hamiltonian in the $l$-bit basis. These coefficients, which decay exponentially in short-range systems \cite{Rademaker+16,Rademaker+17,Thomson+18} and in periodically driven systems~\cite{thomson2020flow}, are strongly modified by the existence of long-range couplings.
In Fig. \ref{fig.pwrhop}, we show these quantities in the case of power-law hopping and nearest-neighbor interactions (corresponding to $\beta=\infty$). The $\tilde{\Delta}_{ij}$ retain their exponentially-decaying nature at short distances, but acquire power-law tails at long range, with a decay exponent $\zeta \approx 2\alpha$ for $\alpha \geq 1$.  This follows immediately from the structure of the eigenstates of the PRBM problem, which are indeed exponentially localized at short distance with power-law tails~\cite{Mirlin+96}.

Secondly, we compute the real-space support of the $l$-bit operators directly. This is something that is extremely natural within the flow equation approach, in contrast to many other numerical methods. Starting from a local density operator $\tilde{n}_i$ defined in the diagonal $l \to \infty$ basis with support only on a single site, we can transform it back into the physical (i.e. real space) basis by inverting the unitary transform used to diagonalize the Hamiltonian. 

The real-space support of the $l$-bits also show power-law tails characteristic of delocalization, after an initial exponential decay at short range. The precise distance where the decay crosses from exponential to power-law depends on the exponent, as well as both the disorder and interaction strength. As $\alpha \to \infty$, the real-space support of the $l$-bits decays exponentially over a larger range before the power-law tail appears, and the resulting $l$-bits closely match the nearest-neighbour case (black dashed line). This further illustrates the critical need for methods able to reach very large system sizes in order to accurately extract the long-distance behaviour of these systems, even in the case of `short-range' ($\alpha > 2d$) power-law exponents.

\begin{figure}[t]
\begin{center}
\includegraphics[width= \linewidth]{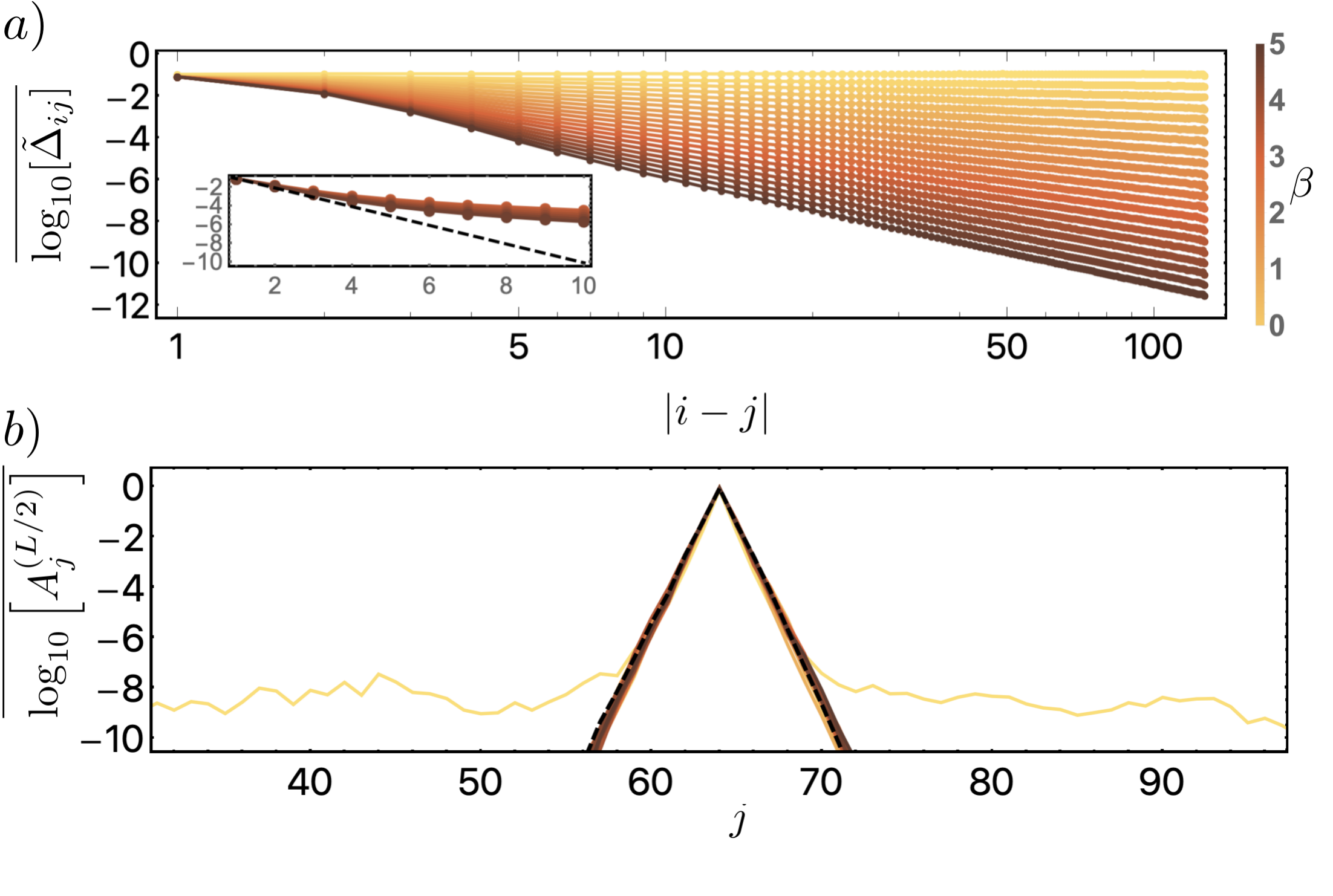}
\caption{$l$-bit interactions (top) and real-space support (bottom) for nearest-neighbor hopping and power-law interactions ($\alpha=\infty$, $\beta \in [0.0,5.0]$ (top to bottom) in increments of $0.25$). a) The disorder-averaged $\tilde{\Delta}_{ij}$  retain their initial power-law distribution for all $\beta$, except at very short distance and large $\beta$ (see inset, semi-log scale, for $\beta \in [3.5,5]$).  b) The $l$-bits remain exponentially localized in real space, with no almost no dependence on $\beta$. The dashed line is the same quantity for a short ranged many-body localized model ($(\alpha \to \infty, \beta \to \infty)$). Chain size $L=128$, disorder realizations $N_s=256$.}
\label{fig.pwrint}
\end{center}
\end{figure}

In Fig. \ref{fig.pwrint}, we show the case of power-law interactions and nearest-neighbor hopping (corresponding to $\alpha=\infty$). The $\tilde{\Delta}_{ij}$ retain their initial power-law distribution at all distances and at all stages during the flow procedure. 
Surprisingly, we find that the real-space support of the $l$-bits is essentially unmodifed by the range of the interactions: they retain their exponentially decaying character even in the limit of $\beta=0$, with only an extremely small extended `tail' appearing following the strong initial exponential decay. This may be an effect of the truncation in Eq. \ref{eqn:ansatz} suppressing degrees of freedom responsible for delocalization, or it may be that delocalization is only seen in higher-order contributions to Eq. \ref{eq.nansatz}, corresponding to multipole processes. 

\subsection{Dynamics of Imbalance and Phase Diagram} 
\label{sec.dyn}
We now move on to study the effect of power-law couplings on the quantum dynamics of the system.  
We set up an initial charge density wave (CDW) state and see how it relaxes under its own quantum dynamics. To monitor this, we define the imbalance as:
\begin{align}
\mathcal{I}(t) = \frac{2}{L} \sum_i (-1)^{i} \langle n_i (t) \rangle
\end{align}
\begin{figure}[t]
\begin{center}
\includegraphics[width= \linewidth]{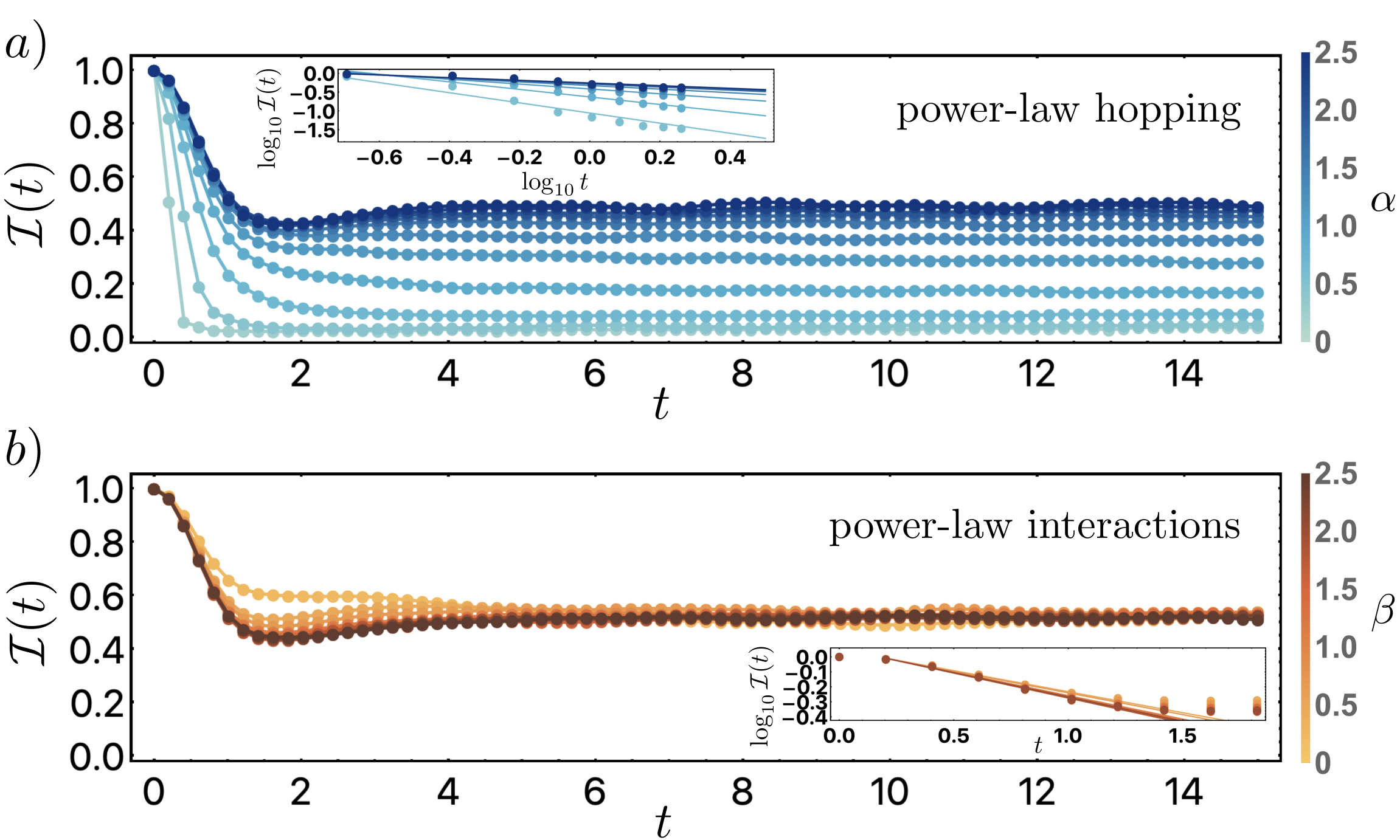}
\caption{Relaxation of the imbalance following a quench from a CDW state with a) power-law hopping $\alpha \in [0.0,2.5]$ in increments of $0.25$ (bottom to top) and $\beta =\infty$ (nearest-neighbour interactions) and b) power-law interactions $\beta \in [0.0,2.5]$ in increments of $0.25$ (top to bottom) with $\alpha = \infty$ (nearest-neighbour hopping). Decreasing $\alpha$ makes the long-time imbalance go to zero (as a power law in time for small $\alpha$, see top inset) whereas changing $\beta$ has almost no effect on the long-time dynamics of the imbalance which approaches a finite plateau almost exponentially (see inset). Chain size $L=64$, disorder realizations $N_s=256$.}
\label{fig.imbalance}
\end{center}
\end{figure}
which involves computing the density dynamics on each lattice site using flow equations, and then summing the results. The long time behavior of the imbalance is often used as a proxy for the MBL transition, since in a localized phase any initial inhomogeneity persists at long time due to enhanced memory of initial conditions while in a thermal, delocalized phase the imbalance is expected to decay to zero as a power law with a disorder-dependent exponent, vanishing at the transition~\cite{LuitzEtALPRB16,BiroliTarziaPRB17}.
Using the time-dependent mean-field decoupling on the effective $l$-bit Hamiltonian, the results for the relaxation dynamics of the imbalance are shown in Figure 3, for chains of length $L=64$ in the cases of power-law hopping with nearest-neighbour interactions (panel a), and nearest-neighbour hopping with power-law interactions (panel b). In Fig.~\ref{fig.imbalance} panel (a), we see that for $\alpha \gtrsim 1$ the system remains localized as for the short-range model, while upon decreasing $\alpha$ the imbalance continuously decrease toward zero, a behavior that is reminscent of the PRBM model and similar models with non-random short-range interactions \cite{Khatami+12}. For $\alpha=0$, the decay of the imbalance is approximately exponential, while for $\alpha>0$ it is consistent with a power-law. On the contrary, Fig.~\ref{fig.imbalance} panel (b) shows that decreasing $\beta$, i.e. making the range of interactions larger, has little to no effect on the long-time imbalance and the system remains localized, with small values of $\beta$ leading to the appearance of a short plateau that vanishes at longer times. Though short-lived, this plateau is intriguing as it suggests that long-range interactions may weakly stabilise localization at short times.

\begin{figure}[t]
\begin{center}
\includegraphics[width= \linewidth]{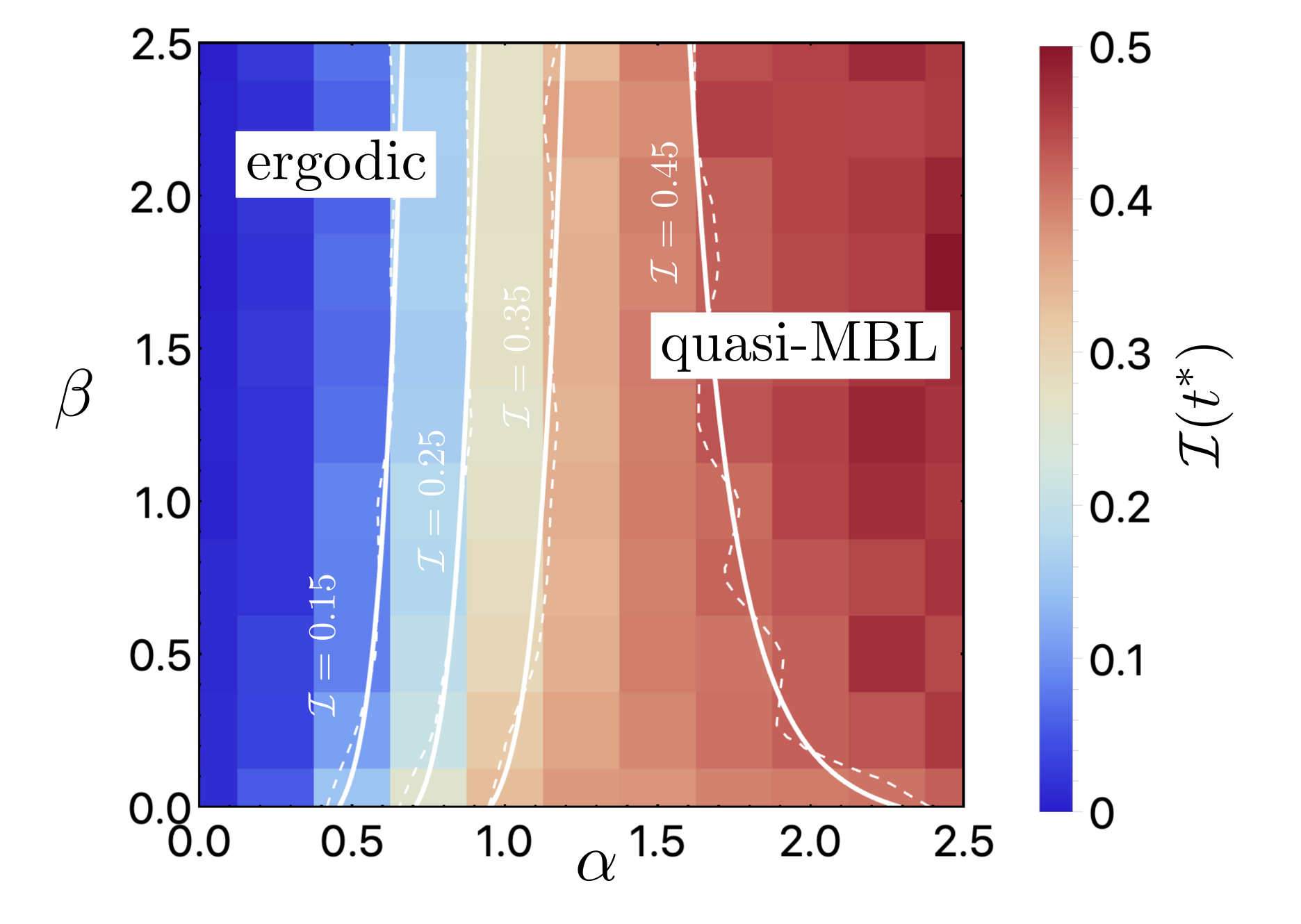}
\caption{Phase diagram of model~(1) as a function of $\alpha$ (hopping exponent) and $\beta$ (interaction exponent).
The colour scale shows the imbalance $\mathcal{I}(t)$ at a time $t^*=100$ following a quench. The dotted lines show contours of the imbalance $\mathcal{I}(t^*)=0.15,0.25,0.35,0.45$ computed using a linear interpolation: the solid white lines are guides to the eye. The system size is $L=64$,  with $50 \leq N_s \leq 128$ disorder realizations, as required for convergence. For $\beta = 0.0,0.25$ and $2.0$, we also took additional data points (not shown) at double the resolution along the $\alpha$ axis in order to ensure that our resolution was sufficient to resolve the main features. }
\label{fig.phase}
\end{center}
\end{figure}
Having examined their effects separately, we now compute the imbalance in the presence of both long-ranged interactions {\it and} long-range hopping, and obtain a qualitative phase diagram shown in Fig. \ref{fig.phase} where we show the imbalance $\mathcal{I}(t)$ at a time $t^*=100$ after the quench as a function of $\alpha,\beta$ and super-impose lines at fixed imbalance as guide to the eye. In the upper-right corner, corresponding to fast decaying hopping and interactions ($\alpha,\beta \geq 2$), the system is in a quasi-MBL phase, with a finite and large imbalance. Keeping $\beta\geq 2$ and decreasing the hopping exponent $\alpha$, the imbalance displays a sharp crossover from localized to delocalized behavior, consistent with the similar model of Ref.~\cite{Khatami+12}. 

We can now ask what happens to those two phases as we increase the range of the interaction, i.e. decreases $\beta$ toward zero. The ergodic phase is expected to be robust to long-range interactions, and indeed we see that the imbalance for $\alpha<1$ remains constant and close to zero upon decreasing $\beta$ (see the almost vertical contour lines) . On the other hand, and quite surprisingly, we find the imbalance to remain strongly unaffected by long range interactions even for $\alpha\gtrsim 1$, consistently with the results of Figure 3 for the $\alpha=\infty$ case. However the lines at fixed imbalance bends towards the right for small $\beta$, suggesting that the localization of the lower right corner of the phase diagram may be less robust than the upper right corner, consistent with a significantly broadened crossover from localised to delocalised behaviour in this regime.

\section{Discussion}  
\label{sec.dis}

\begin{figure}[t]
\begin{center}
\includegraphics[width= \linewidth]{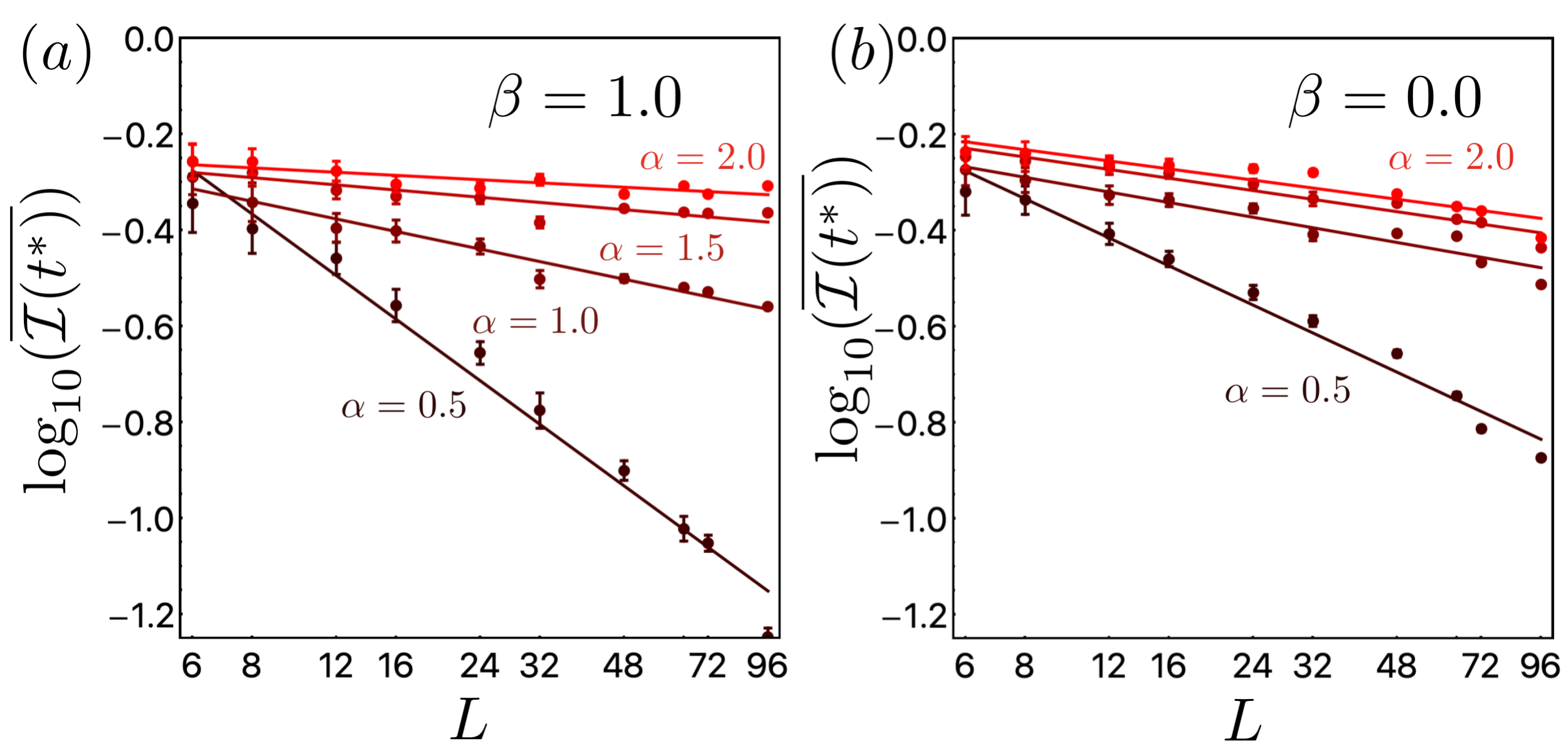}
\caption{Long-time imbalance $\mathcal{I}(t^*)$ (at a time $t^*=10$ following a quench) versus system size $L$ for different values of $(\alpha,\beta)$, averaged over $N_s=256$ disorder realisations for the smallest system sizes ($L=6,8,12, 24, 36$), $N_s=128$ for $L=48$, $N_s=64$ for $L=72$ and $N_s=32$ for $L=96$. The plots are shown on a log-log scale and the solid lines are linear fits to the data. Error bars indicate the variance across disorder realisations.  For $\beta=1.0$ (panel a), the imbalance decreases with system size approximately like a power-law (note the log-log scale) for $\alpha \leq 1$. For $\beta = 0$ (panel b) the imbalance decays with system size for all values of $\alpha$, suggesting slow delocalization with system size. Interestingly, the imbalance in the delocalized (small $\alpha$) regime decays more slowly with system size in the case of long-range interactions ($\beta=0.0$).}
\label{fig.scaling}
\end{center}
\end{figure}

Our results show that upon increasing the range of the hopping, a transition from delocalization to quasi-MBL exists, both for short ranged interactions as well as for $\beta<2$, in a regime where perturbative arguments based on a locator expansion would exclude it.  We have performed extensive checks to validate our approach in this regime, including comparison with exact numerics for small system sizes and monitoring the flow invariant, a sensitive probe of the validity of our scheme. This quasi-MBL phase could also be metastable for finite size and/or finite time: recent works suggest that in the intermediate regime $1 < \beta < 2$  an infinitely large system would be delocalized while finite-size systems will see a localization transition as a function of increasing system size $L$ (or equivalently, exhibit a size-dependent critical disorder $W_c(L)$)~\cite{Burin06,Burin15,Tikhonov+18,Gopalakrishnan+19}.  Our results show (see Appendix B) that the quasi-MBL phase shrinks as the system size is increased, consistent with this argument, and thus we expect that the quasi-MBL phase is likely to be stable for finite-size systems, but unstable in the thermodynamic limit.  To further support this statement we plot in Figure~\ref{fig.scaling} the long-time imbalance $I(t*)$ versus system sizes $L$ for different values of $(\alpha,\beta)$ in the phase diagram. As we can clearly see for $\beta=0$ the imbalance decays like a power-law for all values of $\alpha$  suggesting slow delocalization in the thermodynamic limit. Interestingly, for $\alpha=0.5-1$ the final value of the imbalance is larger for $\beta=0$ than for $\beta=1$, supporting the idea of a broad interaction-induced crossover region that slowly becomes ergodic in the limit of large system sizes. The Gaussian distribution of couplings (with zero mean) could also play a role in the apparent robustness of the localized phase, as by comparison long-range couplings with random signs, as commonly studied in quantum spin models, exhibit enhanced delocalization, shown in Appendix~\ref{app.random_sign}. Finally, it is worth noticing that in the $\alpha,\beta\rightarrow0$ limit, Eq.~(1) reduces to a model of fermions with all-to-all random couplings, reminiscent of the maximally chaotic Sachdev-Ye-Kitaev model~\cite{SachdevYePRL93}. As shown in Ref.~\onlinecite{Garcia+19}, adding finite range hopping to SYK-like models can lead to an increased localized behavior, at least for finite systems, consistent with the results shown here.

On a technical level, there are two key avenues for improving the method further. The first is the incorporation of higher-order terms into the ansatz for both the running Hamiltonian and the running number operator. The necessity of including the normal-ordering corrections makes this procedure extremely algebraically challenging and difficult to automate, however, complicating this procedure significantly. Further work is currently underway on different techniques by which to alleviate this issue. The second route towards improvement is the search for a more optimal generator, perhaps one that does not result in a proliferation of new couplings as the Wegner generator does. Recently, connections between Wegner generators and adiabatic gauge potentials have been noted~\cite{Wurtz+20}, and it is likely that further ongoing work examing this connection will allow systematic improvements to be made to Wegner-type generators, improving their convergence properties and allowing the intelligent design of optimised generators for specific problems, bypassing many of the implementation issues around continuous unitary transforms for arbitrary systems.

\section{Conclusion} 
\label{sec.con}
We have used the flow equation method to study a model of one-dimensional fermions with Gaussian-distributed, power-law decaying hopping and interactions, and diagonal box disorder. For large diagonal disorder, compared to typical scales of interactions and hoppings, we have provided evidence of a transition from a delocalized ergodic phase to a quasi-MBL phase upon increasing the exponent $\alpha$ controlling the range of hopping. A crossover survives even for slowly decaying interactions, $\beta<2$, although it appears to become less sharp. This quasi-MBL phase has intriguing properties such as algebraically decaying $l$-bit interactions. To probe the possible metastability of this phase we studied the decay of long-time imbalance with system sizes, finding signature of slow power-law delocalization, which however appears more effective at finite $\beta$ than in the regime of $\beta\rightarrow0$. Assessing the corresponding lifetime of the quasi-MBL case  as well as the possible existence of a critical disorder strength is an interesting open question for future work. Another open question is the stability of such a phase to the propagation of ergodic bubbles: further investigation based on our model and approach could provide insights into this largely unexplored question, e.g. by studying the coupling of this quasi-MBL phase to an ergodic bath~\cite{CrowleyChandranArxiv19}.

We have also used this work to demonstrate an improved implementation of the truncated flow equation approach, which to date remains the only controlled technique able to compute both the local intergals of motion ($l$-bits) non-perturbatively, as well as to numerically construct the effective Hamiltonian in the $l$-bit basis for large system sizes, particularly in the case of disordered long-range couplings, a situation which is extremely challenging to numerically investigate. We have shown that the method is capable of extremely high accuracy across the entire phase diagram, able to extract both static and dynamic properties, and error estimate both with respect to exact numerical methods and self-consistent quantities remain small for all parameters considered in this work. Our results demonstrate that the truncated flow equation method is an extremely powerful, flexible method for the study of disordered many-body systems, particularly in parameter regimes difficult to acces by other means, and we have shown that it is able to access quantities which are impossible to obtain with other methods. Other recent developments include the extension of flow equation methods to study driven~\cite{thomson2020flow} and dissipative~\cite{rosso2020dissipative} systems, highlighting the versatility and wide applicability of this approach which we hope will become a key numerical method for the study of disordered systems in the near future.

Note: during review, we became aware of another very recent work studying the effect of disordered long-range couplings, the results of which are consistent with those we present here~\cite{prasad2020enhanced}.

\acknowledgements
The computations were performed on the Coll\`ege de France IPH computer cluster. We acknowledge use of the QuSpin exact diagonalisation library for benchmarking our FE code \cite{Weinberg+17,Weinberg+19}, and support from the grants DynDisQ from DIM SIRTEQ and the  ANR  grant  ``NonEQuMat”  (ANR-19-CE47-0001)

\appendix
\section{Normal-ordering}
\label{sec.normal_ordering}
A key ingredient in the calculation is the adoption of a normal-ordering procedure  \cite{Wegner06,Kehrein07,Thomson+18}, which allow us to consistently group together terms at each order of the Hamiltonian, and to incorporate corrections from higher-order terms which are then discarded from our variational manifold. We will assume all contractions will be computed with respect to a product state, and the relevant contractions will be denoted:
\begin{align}
\{c^{\dagger}_i,c_j\} &= G_{ij} + \tilde{G}_{ji} = \delta_{ij} \\
G_{ij} &= \langle c^{\dagger}_i c_j \rangle = \delta_{ij} \langle n_i \rangle\\
\tilde{G}_{ji} &= \langle c_j c^{\dagger}_i \rangle = \delta_{ij} -\langle c^{\dagger}_i c_j \rangle = \delta_{ij} (1 - \langle n_i \rangle)
\end{align}
To calculate the commutators of normal-ordered strings of operators, we need to use the following theorem \cite{Kehrein07}:
\begin{align}
:O_1(A)::O_2(A'): &= :\exp \left( \sum_{ij} G_{ij} \frac{\partial^2}{\partial A_j' \partial A_i}  \right) O_1(A) O_2(A'):
\label{eq.wick}
\end{align}
which, for example, leads to the following commutation relation for pairs of fermion operators:
\begin{align}
& [:c^{\dagger}_{\alpha} c_{\beta}:,:c^{\dagger}_{\gamma} c_{\delta}:] = (G_{\gamma \beta} + \tilde{G}_{\beta \gamma}):c^{\dagger}_{\alpha} c_{\delta}: \nn\\
& \quad \quad \quad - (G_{\alpha \delta} + \tilde{G}_{\delta \alpha}) :c^{\dagger}_{\gamma} c_{\beta}:  \nn\\
& \quad \quad \quad \quad + (G_{\alpha \delta} \tilde{G}_{\beta \gamma} - G){\gamma \beta} \tilde{G}_{\delta \alpha}) \\
&= \delta_{\beta \gamma} :c^{\dagger}_{\alpha} c_{\delta}: - \delta_{\alpha \delta} :c^{\dagger}_{\gamma} c_{\beta}: + (G_{\alpha \delta} \tilde{G}_{\beta \gamma} - G_{\gamma \beta} \tilde{G}_{\delta \alpha}) 
\label{eq.no}
\end{align}
which is just the regular commutator plus a constant. All necessary commutators can be computed from Eq. \ref{eq.wick}, though the calculation is extremely tedious and will not be shown here: for further details, see Refs. \cite{Wegner06,Kehrein07,Thomson+18}. In principle, one should define an $l$-dependent state and recompute the normal-ordering corrections at each flow timestep accordingly, however to capture the main physics it is sufficient to simply pick a target state and compute the corrections with respect to that state \cite{Kehrein07}. In the main text, we compute the contractions with respect to an infinite-temperature product state such that $\langle n_i \rangle = 0.5 \phantom{.} \forall i$. This has the advantage that many of the normal-ordering corrections (e.g. the final terms in Eq. \ref{eq.no} above) vanish identically.

\section{Phase Diagram: Effect of System Size}
\begin{figure}
\begin{center}
\includegraphics[width= \linewidth]{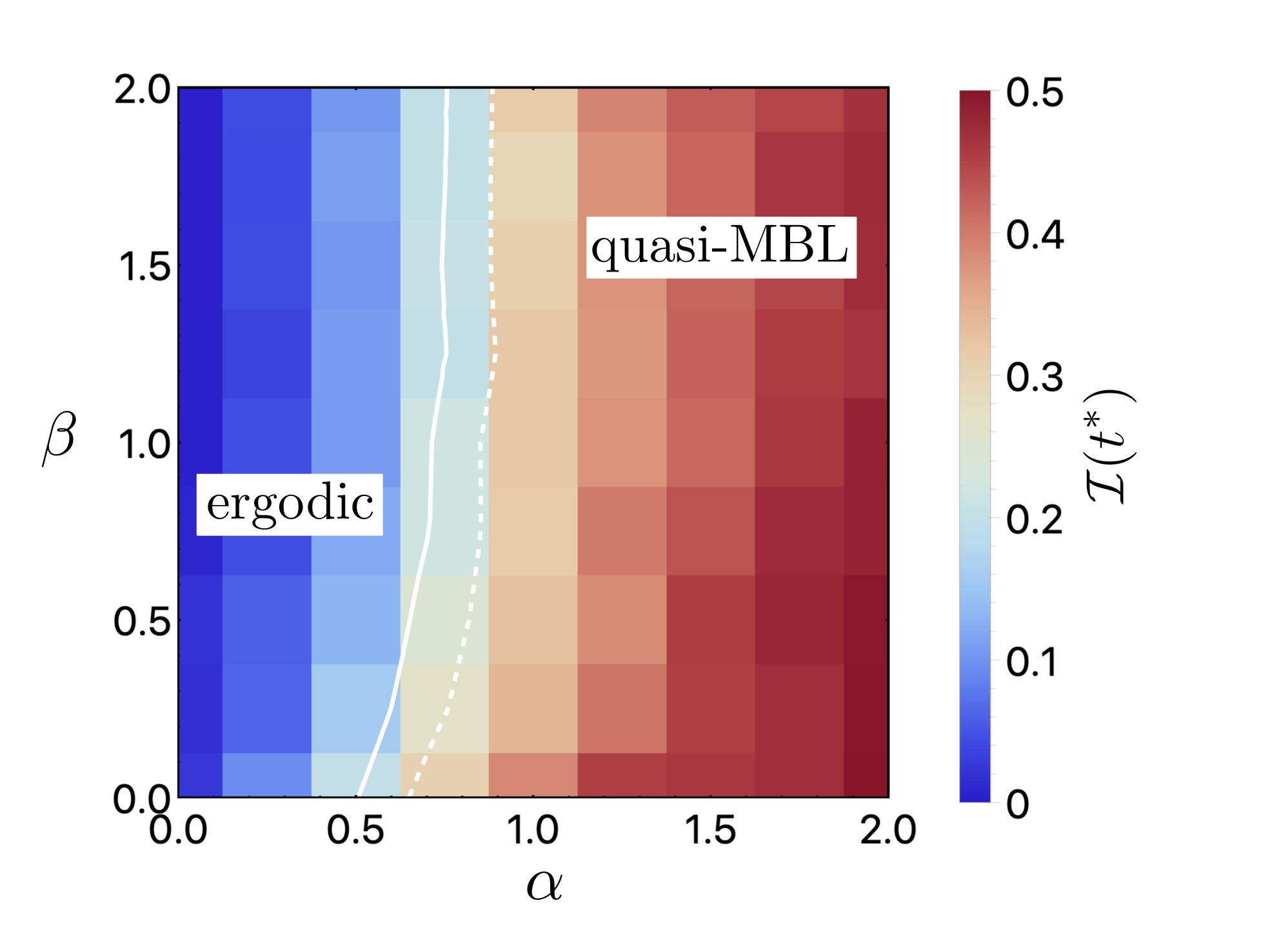}
\caption{The same quantity as in Fig. 4 of the main text, here for system size $L=36$ and averaged over $N_100$ disorder realizations. The solid white line represents $\mathcal{I}(t*=100)=0.25$ (half the maximum value) for the $L=36$ system, and is a rough indicator of the position of the transition, while the dashed white line is the same quantity for the $L=64$ system shown in Fig. 4 of the main text. There is a clear drift of the boundary towards larger values of $\alpha$ as we increase the system size, however the main features are robust. }
\label{fig.phase36}
\end{center}
\end{figure}

To verify our conclusions, we have also computed the phase diagram for a chain of $L=36$ sites averaged over $N_s = 100$ disorder realizations, shown in Fig. \ref{fig.phase36}. The phase boundary moves, as expected, but the general conclusion is the same. This demonstrates that the main features of the phase diagram presented in the main text are robust. The flow invariant remains below a maximum value of $\delta I_2^{max} = 0.012$ at all points in this figure. This data suggests that, all other things being equal, there is a slow growth of the number of resonances as the system size is increased, consistent with the resonance counting arguments in the existing literature. Our results are an indication that even for large system sizes, localization still persists over a large region of the phase diagram. Note however that the reversal of curvature seen in Fig. 4 of the main text  for $\alpha > 1$ is not present in this data, and the $L=36$ system is more localized in this region, with a larger imbalance. This is consistent with the idea that larger systems exhibit more delocalising resonances, destabilising the localized phase.

\section{Random-Sign Disorder}
\label{app.random_sign}
\begin{figure*}[t!]
\begin{center}
\includegraphics[width= \linewidth]{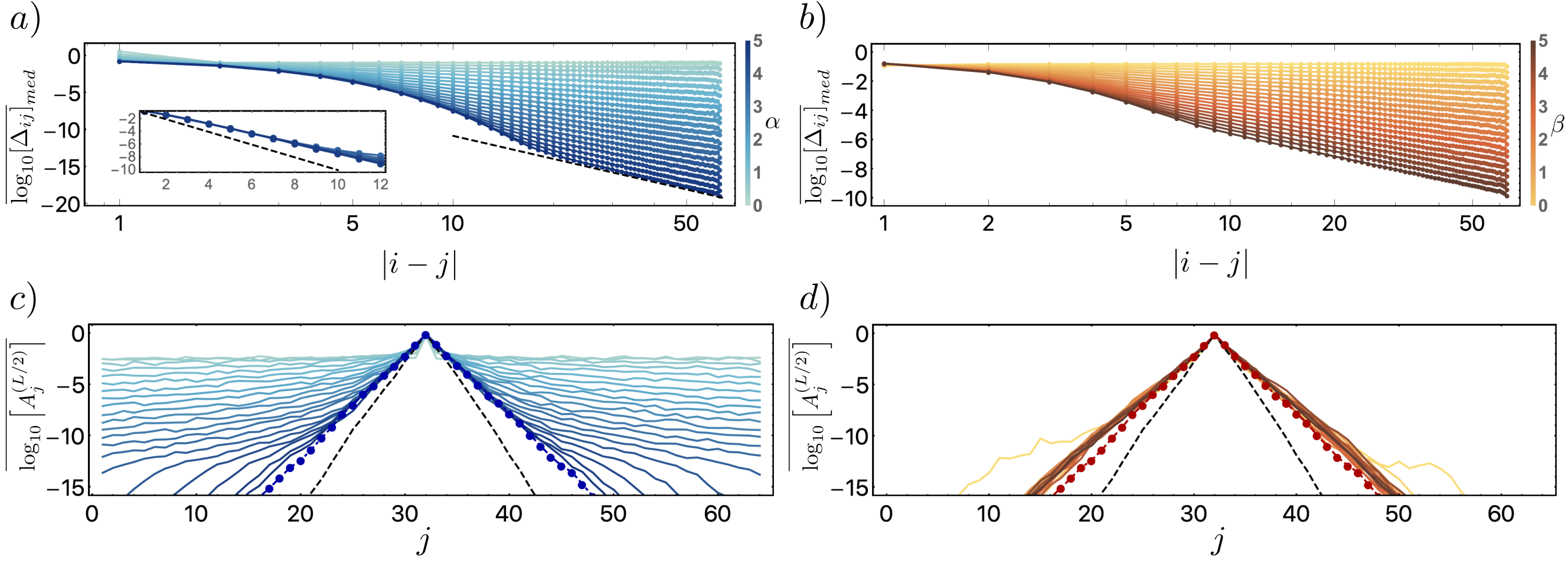}
\caption{Various static properties of the fixed-point Hamiltonian with random-sign disorder, rather than Gaussian-distributed disorder. All data here is taken for system sizes $L=64$ with $N_s=128$ disorder realizations, and the colour schemes are the same as in the main text. The left column shows data for long-range hopping, while the rigth column shows data for long-range interactions. a) Fixed-point couplings $\Delta_{ij}$ in the case of power-law hopping and nearest-neighbour interactions $\beta \to \infty$, again with $\alpha \in [0,5]$ as in the main text. The black dashed lines are the same as in the main text. b) The same quantity plotted for the case of power-law interactions (with $\alpha \to \infty$ and$\beta \in [0,5]$ as before). c) The real-space support of the $l$-bits in the case of long-range hopping. The black dashed line is the same as in the main text (the $\alpha,\beta \to \infty$ limit with Gaussian distributed disorder), while the blue dots show the $\alpha,\beta \to \infty$ limit of the random-sign disorder.  d) The same quantity plotted for long-range interactions, with $\alpha \to \infty$ and $\beta \in [0,5]$. The black dashed line is again the same as in the main text, while the red dots show the $\alpha, \beta \to \infty$ limit of the random-sign disorder.}
\label{fig.rsgn}
\end{center}
\end{figure*}

Previous works on long-range couplings in spin chains have considered so-called `random sign disorder', in which the couplings are fixed in magnitude but allowed to vary in sign, i.e. $J_{ij} = \pm J_0 / |i-j|^{\alpha}$ and $V_{ij} = \pm V_0 / |i-j|^{\beta}$ where the signs are chosen randomly. These works have predicted the absence of a localized phase in the regime $d \leq \beta \leq 2d$, whereas we find clear signs of localization in this regime. While this could be a finite-size effect, or equivalently we may simply be below the critical disorder threshold for this system size, we have nonetheless simulated this type of disorder as well in order to compare with our (zero mean) Gaussian-distributed random couplings. The results are shown in Fig. \ref{fig.rsgn}.

Remarkably, we find that the case of Gaussian-distributed random couplings is indeed significantly more localized than the random-sign disorder, both quantitatively and qualitatively. This effect is most prominent at short distances, with the long-distances tails behaving the same regardless of the specific type of disorder. This difference, while striking at first sight, can be explained simply by the typical magnitude of the coupling terms being large (and, crucially, non-zero) in the case of random-sign disorder, while the typical value is zero for the Gaussian-distributed disorder considered in the main text. This clearly demonstrates that the short-range behaviour of these systems is a complex function of the disorder and the long-range of the couplings, whereas at large distances only the asymptotic form of the disorder is important.

\bibliography{refs}

\begin{thebibliography}{82}
\expandafter\ifx\csname natexlab\endcsname\relax\def\natexlab#1{#1}\fi
\expandafter\ifx\csname bibnamefont\endcsname\relax
  \def\bibnamefont#1{#1}\fi
\expandafter\ifx\csname bibfnamefont\endcsname\relax
  \def\bibfnamefont#1{#1}\fi
\expandafter\ifx\csname citenamefont\endcsname\relax
  \def\citenamefont#1{#1}\fi
\expandafter\ifx\csname url\endcsname\relax
  \def\url#1{\texttt{#1}}\fi
\expandafter\ifx\csname urlprefix\endcsname\relax\def\urlprefix{URL }\fi
\providecommand{\bibinfo}[2]{#2}
\providecommand{\eprint}[2][]{\url{#2}}

\bibitem[{\citenamefont{Rigol et~al.}(2008)\citenamefont{Rigol, Dunjko, and
  Olshanii}}]{RigolETH08}
\bibinfo{author}{\bibfnamefont{M.}~\bibnamefont{Rigol}},
  \bibinfo{author}{\bibfnamefont{V.}~\bibnamefont{Dunjko}}, \bibnamefont{and}
  \bibinfo{author}{\bibfnamefont{M.}~\bibnamefont{Olshanii}},
  \bibinfo{journal}{Nature} \textbf{\bibinfo{volume}{452}},
  \bibinfo{pages}{854} (\bibinfo{year}{2008}).

\bibitem[{\citenamefont{D'Alessio et~al.}(2016)\citenamefont{D'Alessio, Kafri,
  Polkovnikov, and Rigol}}]{DAlessioEtAl2016}
\bibinfo{author}{\bibfnamefont{L.}~\bibnamefont{D'Alessio}},
  \bibinfo{author}{\bibfnamefont{Y.}~\bibnamefont{Kafri}},
  \bibinfo{author}{\bibfnamefont{A.}~\bibnamefont{Polkovnikov}},
  \bibnamefont{and} \bibinfo{author}{\bibfnamefont{M.}~\bibnamefont{Rigol}},
  \bibinfo{journal}{Advances in Physics} \textbf{\bibinfo{volume}{65}},
  \bibinfo{pages}{239} (\bibinfo{year}{2016}).

\bibitem[{\citenamefont{Nandkishore and Huse}(2015)}]{Nandkishore+15}
\bibinfo{author}{\bibfnamefont{R.}~\bibnamefont{Nandkishore}} \bibnamefont{and}
  \bibinfo{author}{\bibfnamefont{D.~A.} \bibnamefont{Huse}},
  \bibinfo{journal}{Annu. Rev. Condens. Matter Phys.}
  \textbf{\bibinfo{volume}{6}}, \bibinfo{pages}{15} (\bibinfo{year}{2015}).

\bibitem[{\citenamefont{Abanin et~al.}(2019)\citenamefont{Abanin, Altman,
  Bloch, and Serbyn}}]{AbaninEtAlRMP19}
\bibinfo{author}{\bibfnamefont{D.~A.} \bibnamefont{Abanin}},
  \bibinfo{author}{\bibfnamefont{E.}~\bibnamefont{Altman}},
  \bibinfo{author}{\bibfnamefont{I.}~\bibnamefont{Bloch}}, \bibnamefont{and}
  \bibinfo{author}{\bibfnamefont{M.}~\bibnamefont{Serbyn}},
  \bibinfo{journal}{Rev. Mod. Phys.} \textbf{\bibinfo{volume}{91}},
  \bibinfo{pages}{021001} (\bibinfo{year}{2019}).

\bibitem[{\citenamefont{Deutsch}(1991)}]{DeutschPRA91}
\bibinfo{author}{\bibfnamefont{J.~M.} \bibnamefont{Deutsch}},
  \bibinfo{journal}{Phys. Rev. A} \textbf{\bibinfo{volume}{43}},
  \bibinfo{pages}{2046} (\bibinfo{year}{1991}).

\bibitem[{\citenamefont{Altman}(2018)}]{AltmanNaturePhysics2018}
\bibinfo{author}{\bibfnamefont{E.}~\bibnamefont{Altman}},
  \bibinfo{journal}{Nature Physics} \textbf{\bibinfo{volume}{14}},
  \bibinfo{pages}{979} (\bibinfo{year}{2018}).

\bibitem[{\citenamefont{Qi}(2018)}]{QiXLNatPhys2018}
\bibinfo{author}{\bibfnamefont{X.-L.} \bibnamefont{Qi}},
  \bibinfo{journal}{Nature Physics} \textbf{\bibinfo{volume}{14}},
  \bibinfo{pages}{984} (\bibinfo{year}{2018}).

\bibitem[{\citenamefont{Dymarsky and Pavlenko}(2019)}]{DymarskyPavlenkoPRL19}
\bibinfo{author}{\bibfnamefont{A.}~\bibnamefont{Dymarsky}} \bibnamefont{and}
  \bibinfo{author}{\bibfnamefont{K.}~\bibnamefont{Pavlenko}},
  \bibinfo{journal}{Phys. Rev. Lett.} \textbf{\bibinfo{volume}{123}},
  \bibinfo{pages}{111602} (\bibinfo{year}{2019}).

\bibitem[{\citenamefont{Huse et~al.}(2013)\citenamefont{Huse, Nandkishore,
  Oganesyan, Pal, and Sondhi}}]{Huse+13}
\bibinfo{author}{\bibfnamefont{D.~A.} \bibnamefont{Huse}},
  \bibinfo{author}{\bibfnamefont{R.}~\bibnamefont{Nandkishore}},
  \bibinfo{author}{\bibfnamefont{V.}~\bibnamefont{Oganesyan}},
  \bibinfo{author}{\bibfnamefont{A.}~\bibnamefont{Pal}}, \bibnamefont{and}
  \bibinfo{author}{\bibfnamefont{S.~L.} \bibnamefont{Sondhi}},
  \bibinfo{journal}{Phys. Rev. B} \textbf{\bibinfo{volume}{88}},
  \bibinfo{pages}{014206} (\bibinfo{year}{2013}).

\bibitem[{\citenamefont{Bahri et~al.}(2015)\citenamefont{Bahri, Vosk, Altman,
  and Vishwanath}}]{Bahri+15}
\bibinfo{author}{\bibfnamefont{Y.}~\bibnamefont{Bahri}},
  \bibinfo{author}{\bibfnamefont{R.}~\bibnamefont{Vosk}},
  \bibinfo{author}{\bibfnamefont{E.}~\bibnamefont{Altman}}, \bibnamefont{and}
  \bibinfo{author}{\bibfnamefont{A.}~\bibnamefont{Vishwanath}},
  \bibinfo{journal}{Nature communications} \textbf{\bibinfo{volume}{6}},
  \bibinfo{pages}{7341} (\bibinfo{year}{2015}).

\bibitem[{\citenamefont{Anderson}(1958)}]{Anderson58}
\bibinfo{author}{\bibfnamefont{P.~W.} \bibnamefont{Anderson}},
  \bibinfo{journal}{Physical review} \textbf{\bibinfo{volume}{109}},
  \bibinfo{pages}{1492} (\bibinfo{year}{1958}).

\bibitem[{\citenamefont{Gornyi et~al.}(2005)\citenamefont{Gornyi, Mirlin, and
  Polyakov}}]{Gornyi+05}
\bibinfo{author}{\bibfnamefont{I.~V.} \bibnamefont{Gornyi}},
  \bibinfo{author}{\bibfnamefont{A.~D.} \bibnamefont{Mirlin}},
  \bibnamefont{and} \bibinfo{author}{\bibfnamefont{D.~G.}
  \bibnamefont{Polyakov}}, \bibinfo{journal}{Phys. Rev. Lett.}
  \textbf{\bibinfo{volume}{95}}, \bibinfo{pages}{206603}
  (\bibinfo{year}{2005}).

\bibitem[{\citenamefont{Basko et~al.}(2006)\citenamefont{Basko, Aleiner, and
  Altshuler}}]{Basko+06}
\bibinfo{author}{\bibfnamefont{D.~M.} \bibnamefont{Basko}},
  \bibinfo{author}{\bibfnamefont{I.~L.} \bibnamefont{Aleiner}},
  \bibnamefont{and} \bibinfo{author}{\bibfnamefont{B.~L.}
  \bibnamefont{Altshuler}}, \bibinfo{journal}{Annals of physics}
  \textbf{\bibinfo{volume}{321}}, \bibinfo{pages}{1126} (\bibinfo{year}{2006}).

\bibitem[{\citenamefont{Pal and Huse}(2010)}]{Pal+10}
\bibinfo{author}{\bibfnamefont{A.}~\bibnamefont{Pal}} \bibnamefont{and}
  \bibinfo{author}{\bibfnamefont{D.~A.} \bibnamefont{Huse}},
  \bibinfo{journal}{Phys. Rev. B} \textbf{\bibinfo{volume}{82}},
  \bibinfo{pages}{174411} (\bibinfo{year}{2010}).

\bibitem[{\citenamefont{Alet and Laflorencie}(2018)}]{Alet+18}
\bibinfo{author}{\bibfnamefont{F.}~\bibnamefont{Alet}} \bibnamefont{and}
  \bibinfo{author}{\bibfnamefont{N.}~\bibnamefont{Laflorencie}},
  \bibinfo{journal}{Comptes Rendus Physique} \textbf{\bibinfo{volume}{19}},
  \bibinfo{pages}{498} (\bibinfo{year}{2018}).

\bibitem[{\citenamefont{Schreiber et~al.}(2015)\citenamefont{Schreiber,
  Hodgman, Bordia, Lüschen, Fischer, Vosk, Altman, Schneider, and
  Bloch}}]{BlochMBL2015}
\bibinfo{author}{\bibfnamefont{M.}~\bibnamefont{Schreiber}},
  \bibinfo{author}{\bibfnamefont{S.~S.} \bibnamefont{Hodgman}},
  \bibinfo{author}{\bibfnamefont{P.}~\bibnamefont{Bordia}},
  \bibinfo{author}{\bibfnamefont{H.~P.} \bibnamefont{Lüschen}},
  \bibinfo{author}{\bibfnamefont{M.~H.} \bibnamefont{Fischer}},
  \bibinfo{author}{\bibfnamefont{R.}~\bibnamefont{Vosk}},
  \bibinfo{author}{\bibfnamefont{E.}~\bibnamefont{Altman}},
  \bibinfo{author}{\bibfnamefont{U.}~\bibnamefont{Schneider}},
  \bibnamefont{and} \bibinfo{author}{\bibfnamefont{I.}~\bibnamefont{Bloch}},
  \bibinfo{journal}{Science} \textbf{\bibinfo{volume}{349}},
  \bibinfo{pages}{842} (\bibinfo{year}{2015}).

\bibitem[{\citenamefont{Kondov et~al.}(2015)\citenamefont{Kondov, McGehee, Xu,
  and DeMarco}}]{Kondov+15}
\bibinfo{author}{\bibfnamefont{S.~S.} \bibnamefont{Kondov}},
  \bibinfo{author}{\bibfnamefont{W.~R.} \bibnamefont{McGehee}},
  \bibinfo{author}{\bibfnamefont{W.}~\bibnamefont{Xu}}, \bibnamefont{and}
  \bibinfo{author}{\bibfnamefont{B.}~\bibnamefont{DeMarco}},
  \bibinfo{journal}{Phys. Rev. Lett.} \textbf{\bibinfo{volume}{114}},
  \bibinfo{pages}{083002} (\bibinfo{year}{2015}).

\bibitem[{\citenamefont{Choi et~al.}(2016)\citenamefont{Choi, Hild, Zeiher,
  Schau{\ss}, Rubio-Abadal, Yefsah, Khemani, Huse, Bloch, and Gross}}]{Choi+16}
\bibinfo{author}{\bibfnamefont{J.-y.} \bibnamefont{Choi}},
  \bibinfo{author}{\bibfnamefont{S.}~\bibnamefont{Hild}},
  \bibinfo{author}{\bibfnamefont{J.}~\bibnamefont{Zeiher}},
  \bibinfo{author}{\bibfnamefont{P.}~\bibnamefont{Schau{\ss}}},
  \bibinfo{author}{\bibfnamefont{A.}~\bibnamefont{Rubio-Abadal}},
  \bibinfo{author}{\bibfnamefont{T.}~\bibnamefont{Yefsah}},
  \bibinfo{author}{\bibfnamefont{V.}~\bibnamefont{Khemani}},
  \bibinfo{author}{\bibfnamefont{D.~A.} \bibnamefont{Huse}},
  \bibinfo{author}{\bibfnamefont{I.}~\bibnamefont{Bloch}}, \bibnamefont{and}
  \bibinfo{author}{\bibfnamefont{C.}~\bibnamefont{Gross}},
  \bibinfo{journal}{Science} \textbf{\bibinfo{volume}{352}},
  \bibinfo{pages}{1547} (\bibinfo{year}{2016}).

\bibitem[{\citenamefont{Rispoli et~al.}(2019)\citenamefont{Rispoli, Lukin,
  Schittko, Kim, Tai, L{\'e}onard, and Greiner}}]{RispoliEtAlNature19}
\bibinfo{author}{\bibfnamefont{M.}~\bibnamefont{Rispoli}},
  \bibinfo{author}{\bibfnamefont{A.}~\bibnamefont{Lukin}},
  \bibinfo{author}{\bibfnamefont{R.}~\bibnamefont{Schittko}},
  \bibinfo{author}{\bibfnamefont{S.}~\bibnamefont{Kim}},
  \bibinfo{author}{\bibfnamefont{M.~E.} \bibnamefont{Tai}},
  \bibinfo{author}{\bibfnamefont{J.}~\bibnamefont{L{\'e}onard}},
  \bibnamefont{and} \bibinfo{author}{\bibfnamefont{M.}~\bibnamefont{Greiner}},
  \bibinfo{journal}{Nature} \textbf{\bibinfo{volume}{573}},
  \bibinfo{pages}{385} (\bibinfo{year}{2019}).

\bibitem[{\citenamefont{Lukin et~al.}(2019)\citenamefont{Lukin, Rispoli,
  Schittko, Tai, Kaufman, Choi, Khemani, L{\'e}onard, and
  Greiner}}]{LukinScience2019}
\bibinfo{author}{\bibfnamefont{A.}~\bibnamefont{Lukin}},
  \bibinfo{author}{\bibfnamefont{M.}~\bibnamefont{Rispoli}},
  \bibinfo{author}{\bibfnamefont{R.}~\bibnamefont{Schittko}},
  \bibinfo{author}{\bibfnamefont{M.~E.} \bibnamefont{Tai}},
  \bibinfo{author}{\bibfnamefont{A.~M.} \bibnamefont{Kaufman}},
  \bibinfo{author}{\bibfnamefont{S.}~\bibnamefont{Choi}},
  \bibinfo{author}{\bibfnamefont{V.}~\bibnamefont{Khemani}},
  \bibinfo{author}{\bibfnamefont{J.}~\bibnamefont{L{\'e}onard}},
  \bibnamefont{and} \bibinfo{author}{\bibfnamefont{M.}~\bibnamefont{Greiner}},
  \bibinfo{journal}{Science} \textbf{\bibinfo{volume}{364}},
  \bibinfo{pages}{256} (\bibinfo{year}{2019}), ISSN \bibinfo{issn}{0036-8075}.

\bibitem[{\citenamefont{Smith et~al.}(2016)\citenamefont{Smith, Lee, Richerme,
  Neyenhuis, Hess, Hauke, Heyl, Huse, and Monroe}}]{Smith+16}
\bibinfo{author}{\bibfnamefont{J.}~\bibnamefont{Smith}},
  \bibinfo{author}{\bibfnamefont{A.}~\bibnamefont{Lee}},
  \bibinfo{author}{\bibfnamefont{P.}~\bibnamefont{Richerme}},
  \bibinfo{author}{\bibfnamefont{B.}~\bibnamefont{Neyenhuis}},
  \bibinfo{author}{\bibfnamefont{P.~W.} \bibnamefont{Hess}},
  \bibinfo{author}{\bibfnamefont{P.}~\bibnamefont{Hauke}},
  \bibinfo{author}{\bibfnamefont{M.}~\bibnamefont{Heyl}},
  \bibinfo{author}{\bibfnamefont{D.~A.} \bibnamefont{Huse}}, \bibnamefont{and}
  \bibinfo{author}{\bibfnamefont{C.}~\bibnamefont{Monroe}},
  \bibinfo{journal}{Nature Physics} \textbf{\bibinfo{volume}{12}},
  \bibinfo{pages}{907} (\bibinfo{year}{2016}).

\bibitem[{\citenamefont{Zhang et~al.}(2017)\citenamefont{Zhang, Hess,
  Kyprianidis, Becker, Lee, Smith, Pagano, Potirniche, Potter, Vishwanath
  et~al.}}]{Zhang+17}
\bibinfo{author}{\bibfnamefont{J.}~\bibnamefont{Zhang}},
  \bibinfo{author}{\bibfnamefont{P.~W.} \bibnamefont{Hess}},
  \bibinfo{author}{\bibfnamefont{A.}~\bibnamefont{Kyprianidis}},
  \bibinfo{author}{\bibfnamefont{P.}~\bibnamefont{Becker}},
  \bibinfo{author}{\bibfnamefont{A.}~\bibnamefont{Lee}},
  \bibinfo{author}{\bibfnamefont{J.}~\bibnamefont{Smith}},
  \bibinfo{author}{\bibfnamefont{G.}~\bibnamefont{Pagano}},
  \bibinfo{author}{\bibfnamefont{I.~D.} \bibnamefont{Potirniche}},
  \bibinfo{author}{\bibfnamefont{A.~C.} \bibnamefont{Potter}},
  \bibinfo{author}{\bibfnamefont{A.}~\bibnamefont{Vishwanath}},
  \bibnamefont{et~al.}, \bibinfo{journal}{Nature}
  \textbf{\bibinfo{volume}{543}}, \bibinfo{pages}{217} (\bibinfo{year}{2017}).

\bibitem[{\citenamefont{{\'A}lvarez et~al.}(2015)\citenamefont{{\'A}lvarez,
  Suter, and Kaiser}}]{Alvarez846}
\bibinfo{author}{\bibfnamefont{G.~A.} \bibnamefont{{\'A}lvarez}},
  \bibinfo{author}{\bibfnamefont{D.}~\bibnamefont{Suter}}, \bibnamefont{and}
  \bibinfo{author}{\bibfnamefont{R.}~\bibnamefont{Kaiser}},
  \bibinfo{journal}{Science} \textbf{\bibinfo{volume}{349}},
  \bibinfo{pages}{846} (\bibinfo{year}{2015}), ISSN \bibinfo{issn}{0036-8075}.

\bibitem[{\citenamefont{Kucsko et~al.}(2018)\citenamefont{Kucsko, Choi, Choi,
  Maurer, Zhou, Landig, Sumiya, Onoda, Isoya, Jelezko
  et~al.}}]{KucskoEtAlPRL18}
\bibinfo{author}{\bibfnamefont{G.}~\bibnamefont{Kucsko}},
  \bibinfo{author}{\bibfnamefont{S.}~\bibnamefont{Choi}},
  \bibinfo{author}{\bibfnamefont{J.}~\bibnamefont{Choi}},
  \bibinfo{author}{\bibfnamefont{P.~C.} \bibnamefont{Maurer}},
  \bibinfo{author}{\bibfnamefont{H.}~\bibnamefont{Zhou}},
  \bibinfo{author}{\bibfnamefont{R.}~\bibnamefont{Landig}},
  \bibinfo{author}{\bibfnamefont{H.}~\bibnamefont{Sumiya}},
  \bibinfo{author}{\bibfnamefont{S.}~\bibnamefont{Onoda}},
  \bibinfo{author}{\bibfnamefont{J.}~\bibnamefont{Isoya}},
  \bibinfo{author}{\bibfnamefont{F.}~\bibnamefont{Jelezko}},
  \bibnamefont{et~al.}, \bibinfo{journal}{Phys. Rev. Lett.}
  \textbf{\bibinfo{volume}{121}}, \bibinfo{pages}{023601}
  (\bibinfo{year}{2018}).

\bibitem[{\citenamefont{Serbyn et~al.}(2013)\citenamefont{Serbyn,
  Papi\ifmmode~\acute{c}\else \'{c}\fi{}, and
  Abanin}}]{SerbynPapicAbaninPRL13_2}
\bibinfo{author}{\bibfnamefont{M.}~\bibnamefont{Serbyn}},
  \bibinfo{author}{\bibfnamefont{Z.}~\bibnamefont{Papi\ifmmode~\acute{c}\else
  \'{c}\fi{}}}, \bibnamefont{and} \bibinfo{author}{\bibfnamefont{D.~A.}
  \bibnamefont{Abanin}}, \bibinfo{journal}{Phys. Rev. Lett.}
  \textbf{\bibinfo{volume}{111}}, \bibinfo{pages}{127201}
  (\bibinfo{year}{2013}).

\bibitem[{\citenamefont{Huse et~al.}(2014)\citenamefont{Huse, Nandkishore, and
  Oganesyan}}]{HuseNandkishoreOganesyanPRB14}
\bibinfo{author}{\bibfnamefont{D.~A.} \bibnamefont{Huse}},
  \bibinfo{author}{\bibfnamefont{R.}~\bibnamefont{Nandkishore}},
  \bibnamefont{and}
  \bibinfo{author}{\bibfnamefont{V.}~\bibnamefont{Oganesyan}},
  \bibinfo{journal}{Phys. Rev. B} \textbf{\bibinfo{volume}{90}},
  \bibinfo{pages}{174202} (\bibinfo{year}{2014}).

\bibitem[{\citenamefont{Imbrie}(2016)}]{Imbrie16a}
\bibinfo{author}{\bibfnamefont{J.~Z.} \bibnamefont{Imbrie}},
  \bibinfo{journal}{Journal of Statistical Physics}
  \textbf{\bibinfo{volume}{163}}, \bibinfo{pages}{998} (\bibinfo{year}{2016}),
  ISSN \bibinfo{issn}{1572-9613}.

\bibitem[{\citenamefont{Imbrie et~al.}(2017)\citenamefont{Imbrie, Ros, and
  Scardicchio}}]{Imbrie+16b}
\bibinfo{author}{\bibfnamefont{J.~Z.} \bibnamefont{Imbrie}},
  \bibinfo{author}{\bibfnamefont{V.}~\bibnamefont{Ros}}, \bibnamefont{and}
  \bibinfo{author}{\bibfnamefont{A.}~\bibnamefont{Scardicchio}},
  \bibinfo{journal}{Annalen der Physik} \textbf{\bibinfo{volume}{529}},
  \bibinfo{pages}{1600278} (\bibinfo{year}{2017}).

\bibitem[{\citenamefont{Burin}(2006)}]{Burin06}
\bibinfo{author}{\bibfnamefont{A.~L.} \bibnamefont{Burin}}
  (\bibinfo{year}{2006}), \eprint{cond-mat/0611387}.

\bibitem[{\citenamefont{Yao et~al.}(2014)\citenamefont{Yao, Laumann,
  Gopalakrishnan, Knap, Mueller, Demler, and Lukin}}]{Yao+14}
\bibinfo{author}{\bibfnamefont{N.~Y.} \bibnamefont{Yao}},
  \bibinfo{author}{\bibfnamefont{C.~R.} \bibnamefont{Laumann}},
  \bibinfo{author}{\bibfnamefont{S.}~\bibnamefont{Gopalakrishnan}},
  \bibinfo{author}{\bibfnamefont{M.}~\bibnamefont{Knap}},
  \bibinfo{author}{\bibfnamefont{M.}~\bibnamefont{Mueller}},
  \bibinfo{author}{\bibfnamefont{E.~A.} \bibnamefont{Demler}},
  \bibnamefont{and} \bibinfo{author}{\bibfnamefont{M.~D.} \bibnamefont{Lukin}},
  \bibinfo{journal}{Physical review letters} \textbf{\bibinfo{volume}{113}},
  \bibinfo{pages}{243002} (\bibinfo{year}{2014}).

\bibitem[{\citenamefont{Burin}(2015)}]{Burin15}
\bibinfo{author}{\bibfnamefont{A.~L.} \bibnamefont{Burin}},
  \bibinfo{journal}{Physical Review B} \textbf{\bibinfo{volume}{92}},
  \bibinfo{pages}{104428} (\bibinfo{year}{2015}).

\bibitem[{\citenamefont{Gutman et~al.}(2016)\citenamefont{Gutman, Protopopov,
  Burin, Gornyi, Santos, and Mirlin}}]{Gutman+16}
\bibinfo{author}{\bibfnamefont{D.~B.} \bibnamefont{Gutman}},
  \bibinfo{author}{\bibfnamefont{I.~V.} \bibnamefont{Protopopov}},
  \bibinfo{author}{\bibfnamefont{A.~L.} \bibnamefont{Burin}},
  \bibinfo{author}{\bibfnamefont{I.~V.} \bibnamefont{Gornyi}},
  \bibinfo{author}{\bibfnamefont{R.~A.} \bibnamefont{Santos}},
  \bibnamefont{and} \bibinfo{author}{\bibfnamefont{A.~D.}
  \bibnamefont{Mirlin}}, \bibinfo{journal}{Physical Review B}
  \textbf{\bibinfo{volume}{93}}, \bibinfo{pages}{245427}
  (\bibinfo{year}{2016}).

\bibitem[{\citenamefont{De~Roeck and Huveneers}(2017)}]{DeRoeckHuveneersPRB17}
\bibinfo{author}{\bibfnamefont{W.}~\bibnamefont{De~Roeck}} \bibnamefont{and}
  \bibinfo{author}{\bibfnamefont{F.}~\bibnamefont{Huveneers}},
  \bibinfo{journal}{Phys. Rev. B} \textbf{\bibinfo{volume}{95}},
  \bibinfo{pages}{155129} (\bibinfo{year}{2017}).

\bibitem[{\citenamefont{Kloss and Lev}(2019)}]{Kloss+19}
\bibinfo{author}{\bibfnamefont{B.}~\bibnamefont{Kloss}} \bibnamefont{and}
  \bibinfo{author}{\bibfnamefont{Y.~B.} \bibnamefont{Lev}},
  \bibinfo{journal}{arXiv preprint arXiv:1911.07857}  (\bibinfo{year}{2019}).

\bibitem[{\citenamefont{Nandkishore and Sondhi}(2017)}]{Nandkishore+17}
\bibinfo{author}{\bibfnamefont{R.~M.} \bibnamefont{Nandkishore}}
  \bibnamefont{and} \bibinfo{author}{\bibfnamefont{S.~L.}
  \bibnamefont{Sondhi}}, \bibinfo{journal}{Physical Review X}
  \textbf{\bibinfo{volume}{7}}, \bibinfo{pages}{041021} (\bibinfo{year}{2017}).

\bibitem[{\citenamefont{Santos et~al.}(2016)\citenamefont{Santos, Borgonovi,
  and Celardo}}]{Santos+16}
\bibinfo{author}{\bibfnamefont{L.~F.} \bibnamefont{Santos}},
  \bibinfo{author}{\bibfnamefont{F.}~\bibnamefont{Borgonovi}},
  \bibnamefont{and} \bibinfo{author}{\bibfnamefont{G.~L.}
  \bibnamefont{Celardo}}, \bibinfo{journal}{Physical review letters}
  \textbf{\bibinfo{volume}{116}}, \bibinfo{pages}{250402}
  (\bibinfo{year}{2016}).

\bibitem[{\citenamefont{Roy and Logan}(2019)}]{Roy+19}
\bibinfo{author}{\bibfnamefont{S.}~\bibnamefont{Roy}} \bibnamefont{and}
  \bibinfo{author}{\bibfnamefont{D.~E.} \bibnamefont{Logan}},
  \bibinfo{journal}{SciPost Phys.} \textbf{\bibinfo{volume}{7}},
  \bibinfo{pages}{42} (\bibinfo{year}{2019}).

\bibitem[{\citenamefont{Deng et~al.}(2019)\citenamefont{Deng, Masella, Pupillo,
  and Santos}}]{DengEtAlArxiv19}
\bibinfo{author}{\bibfnamefont{X.}~\bibnamefont{Deng}},
  \bibinfo{author}{\bibfnamefont{G.}~\bibnamefont{Masella}},
  \bibinfo{author}{\bibfnamefont{G.}~\bibnamefont{Pupillo}}, \bibnamefont{and}
  \bibinfo{author}{\bibfnamefont{L.}~\bibnamefont{Santos}}
  (\bibinfo{year}{2019}), \eprint{1912.08131}.

\bibitem[{\citenamefont{Deng et~al.}(2018)\citenamefont{Deng, Kravtsov,
  Shlyapnikov, and Santos}}]{DengEtAlPRL18}
\bibinfo{author}{\bibfnamefont{X.}~\bibnamefont{Deng}},
  \bibinfo{author}{\bibfnamefont{V.~E.} \bibnamefont{Kravtsov}},
  \bibinfo{author}{\bibfnamefont{G.~V.} \bibnamefont{Shlyapnikov}},
  \bibnamefont{and} \bibinfo{author}{\bibfnamefont{L.}~\bibnamefont{Santos}},
  \bibinfo{journal}{Phys. Rev. Lett.} \textbf{\bibinfo{volume}{120}},
  \bibinfo{pages}{110602} (\bibinfo{year}{2018}).

\bibitem[{\citenamefont{Nosov et~al.}(2019)\citenamefont{Nosov, Khaymovich, and
  Kravtsov}}]{NosovEtAlPRB19}
\bibinfo{author}{\bibfnamefont{P.~A.} \bibnamefont{Nosov}},
  \bibinfo{author}{\bibfnamefont{I.~M.} \bibnamefont{Khaymovich}},
  \bibnamefont{and} \bibinfo{author}{\bibfnamefont{V.~E.}
  \bibnamefont{Kravtsov}}, \bibinfo{journal}{Phys. Rev. B}
  \textbf{\bibinfo{volume}{99}}, \bibinfo{pages}{104203}
  (\bibinfo{year}{2019}).

\bibitem[{\citenamefont{Thomson and Schir{\'o}}(2018)}]{Thomson+18}
\bibinfo{author}{\bibfnamefont{S.~J.} \bibnamefont{Thomson}} \bibnamefont{and}
  \bibinfo{author}{\bibfnamefont{M.}~\bibnamefont{Schir{\'o}}},
  \bibinfo{journal}{Physical Review B} \textbf{\bibinfo{volume}{97}},
  \bibinfo{pages}{060201(R)} (\bibinfo{year}{2018}).

\bibitem[{\citenamefont{Thomson and Schir\'o}(2020)}]{Thomson+20}
\bibinfo{author}{\bibfnamefont{S.~J.} \bibnamefont{Thomson}} \bibnamefont{and}
  \bibinfo{author}{\bibfnamefont{M.}~\bibnamefont{Schir\'o}},
  \bibinfo{journal}{Eur. Phy. J. B}  (\bibinfo{year}{2020}).

\bibitem[{\citenamefont{De~Tomasi}(2019)}]{DeTomasiPRB19}
\bibinfo{author}{\bibfnamefont{G.}~\bibnamefont{De~Tomasi}},
  \bibinfo{journal}{Phys. Rev. B} \textbf{\bibinfo{volume}{99}},
  \bibinfo{pages}{054204} (\bibinfo{year}{2019}).

\bibitem[{\citenamefont{Yeung and Oono}(1987)}]{Yeung+87}
\bibinfo{author}{\bibfnamefont{C.}~\bibnamefont{Yeung}} \bibnamefont{and}
  \bibinfo{author}{\bibfnamefont{Y.}~\bibnamefont{Oono}}, \bibinfo{journal}{EPL
  (Europhysics Letters)} \textbf{\bibinfo{volume}{4}}, \bibinfo{pages}{1061}
  (\bibinfo{year}{1987}).

\bibitem[{\citenamefont{Levitov}(1990)}]{Levitov90}
\bibinfo{author}{\bibfnamefont{L.~S.} \bibnamefont{Levitov}},
  \bibinfo{journal}{Physical review letters} \textbf{\bibinfo{volume}{64}},
  \bibinfo{pages}{547} (\bibinfo{year}{1990}).

\bibitem[{\citenamefont{Mirlin et~al.}(1996)\citenamefont{Mirlin, Fyodorov,
  Dittes, Quezada, and Seligman}}]{Mirlin+96}
\bibinfo{author}{\bibfnamefont{A.~D.} \bibnamefont{Mirlin}},
  \bibinfo{author}{\bibfnamefont{Y.~V.} \bibnamefont{Fyodorov}},
  \bibinfo{author}{\bibfnamefont{F.-M.} \bibnamefont{Dittes}},
  \bibinfo{author}{\bibfnamefont{J.}~\bibnamefont{Quezada}}, \bibnamefont{and}
  \bibinfo{author}{\bibfnamefont{T.~H.} \bibnamefont{Seligman}},
  \bibinfo{journal}{Physical Review E} \textbf{\bibinfo{volume}{54}},
  \bibinfo{pages}{3221} (\bibinfo{year}{1996}).

\bibitem[{\citenamefont{Levitov}(1999)}]{Levitov99}
\bibinfo{author}{\bibfnamefont{L.}~\bibnamefont{Levitov}},
  \bibinfo{journal}{Annalen der Physik} \textbf{\bibinfo{volume}{8}},
  \bibinfo{pages}{697} (\bibinfo{year}{1999}).

\bibitem[{\citenamefont{Varga and Braun}(2000)}]{Varga+00}
\bibinfo{author}{\bibfnamefont{I.}~\bibnamefont{Varga}} \bibnamefont{and}
  \bibinfo{author}{\bibfnamefont{D.}~\bibnamefont{Braun}},
  \bibinfo{journal}{Physical Review B} \textbf{\bibinfo{volume}{61}},
  \bibinfo{pages}{R11859} (\bibinfo{year}{2000}).

\bibitem[{\citenamefont{Mirlin and Evers}(2000)}]{Mirlin+00}
\bibinfo{author}{\bibfnamefont{A.~D.} \bibnamefont{Mirlin}} \bibnamefont{and}
  \bibinfo{author}{\bibfnamefont{F.}~\bibnamefont{Evers}},
  \bibinfo{journal}{Physical Review B} \textbf{\bibinfo{volume}{62}},
  \bibinfo{pages}{7920} (\bibinfo{year}{2000}).

\bibitem[{\citenamefont{Evers and Mirlin}(2000)}]{Evers+00}
\bibinfo{author}{\bibfnamefont{F.}~\bibnamefont{Evers}} \bibnamefont{and}
  \bibinfo{author}{\bibfnamefont{A.~D.} \bibnamefont{Mirlin}},
  \bibinfo{journal}{Physical Review Letters} \textbf{\bibinfo{volume}{84}},
  \bibinfo{pages}{3690} (\bibinfo{year}{2000}).

\bibitem[{\citenamefont{Kravtsov et~al.}(2006)\citenamefont{Kravtsov,
  Yevtushenko, and Cuevas}}]{Kravtsov+05}
\bibinfo{author}{\bibfnamefont{V.~E.} \bibnamefont{Kravtsov}},
  \bibinfo{author}{\bibfnamefont{O.}~\bibnamefont{Yevtushenko}},
  \bibnamefont{and} \bibinfo{author}{\bibfnamefont{E.}~\bibnamefont{Cuevas}},
  \bibinfo{journal}{Journal of Physics A: Mathematical and General}
  \textbf{\bibinfo{volume}{39}}, \bibinfo{pages}{2021} (\bibinfo{year}{2006}).

\bibitem[{\citenamefont{Evers and Mirlin}(2008)}]{Evers+08}
\bibinfo{author}{\bibfnamefont{F.}~\bibnamefont{Evers}} \bibnamefont{and}
  \bibinfo{author}{\bibfnamefont{A.~D.} \bibnamefont{Mirlin}},
  \bibinfo{journal}{Reviews of Modern Physics} \textbf{\bibinfo{volume}{80}},
  \bibinfo{pages}{1355} (\bibinfo{year}{2008}).

\bibitem[{\citenamefont{Khatami et~al.}(2012)\citenamefont{Khatami, Rigol,
  Relano, and Garc{\'\i}a-Garc{\'\i}a}}]{Khatami+12}
\bibinfo{author}{\bibfnamefont{E.}~\bibnamefont{Khatami}},
  \bibinfo{author}{\bibfnamefont{M.}~\bibnamefont{Rigol}},
  \bibinfo{author}{\bibfnamefont{A.}~\bibnamefont{Relano}}, \bibnamefont{and}
  \bibinfo{author}{\bibfnamefont{A.~M.} \bibnamefont{Garc{\'\i}a-Garc{\'\i}a}},
  \bibinfo{journal}{Physical Review E} \textbf{\bibinfo{volume}{85}},
  \bibinfo{pages}{050102} (\bibinfo{year}{2012}).

\bibitem[{\citenamefont{Sachdev and Ye}(1993)}]{SachdevYePRL93}
\bibinfo{author}{\bibfnamefont{S.}~\bibnamefont{Sachdev}} \bibnamefont{and}
  \bibinfo{author}{\bibfnamefont{J.}~\bibnamefont{Ye}}, \bibinfo{journal}{Phys.
  Rev. Lett.} \textbf{\bibinfo{volume}{70}}, \bibinfo{pages}{3339}
  (\bibinfo{year}{1993}).

\bibitem[{\citenamefont{Tikhonov and Mirlin}(2018)}]{Tikhonov+18}
\bibinfo{author}{\bibfnamefont{K.~S.} \bibnamefont{Tikhonov}} \bibnamefont{and}
  \bibinfo{author}{\bibfnamefont{A.~D.} \bibnamefont{Mirlin}},
  \bibinfo{journal}{Physical Review B} \textbf{\bibinfo{volume}{97}},
  \bibinfo{pages}{214205} (\bibinfo{year}{2018}).

\bibitem[{\citenamefont{Wegner}(1994)}]{Wegner94}
\bibinfo{author}{\bibfnamefont{F.}~\bibnamefont{Wegner}},
  \bibinfo{journal}{Annalen der physik} \textbf{\bibinfo{volume}{506}},
  \bibinfo{pages}{77} (\bibinfo{year}{1994}).

\bibitem[{\citenamefont{Kehrein}(2007)}]{Kehrein07}
\bibinfo{author}{\bibfnamefont{S.}~\bibnamefont{Kehrein}},
  \emph{\bibinfo{title}{The flow equation approach to many-particle systems}},
  vol. \bibinfo{volume}{217} (\bibinfo{publisher}{Springer},
  \bibinfo{year}{2007}).

\bibitem[{\citenamefont{Moeckel and Kehrein}(2008)}]{Moeckel+08}
\bibinfo{author}{\bibfnamefont{M.}~\bibnamefont{Moeckel}} \bibnamefont{and}
  \bibinfo{author}{\bibfnamefont{S.}~\bibnamefont{Kehrein}},
  \bibinfo{journal}{Phys. Rev. Lett.} \textbf{\bibinfo{volume}{100}},
  \bibinfo{pages}{175702} (\bibinfo{year}{2008}).

\bibitem[{\citenamefont{Hackl and Kehrein}(2008)}]{Hackl+08}
\bibinfo{author}{\bibfnamefont{A.}~\bibnamefont{Hackl}} \bibnamefont{and}
  \bibinfo{author}{\bibfnamefont{S.}~\bibnamefont{Kehrein}},
  \bibinfo{journal}{Phys. Rev. B} \textbf{\bibinfo{volume}{78}},
  \bibinfo{pages}{092303} (\bibinfo{year}{2008}).

\bibitem[{\citenamefont{Hackl and Kehrein}(2009)}]{Hackl+09}
\bibinfo{author}{\bibfnamefont{A.}~\bibnamefont{Hackl}} \bibnamefont{and}
  \bibinfo{author}{\bibfnamefont{S.}~\bibnamefont{Kehrein}},
  \bibinfo{journal}{Journal of Physics: Condensed Matter}
  \textbf{\bibinfo{volume}{21}}, \bibinfo{pages}{015601}
  (\bibinfo{year}{2009}).

\bibitem[{\citenamefont{Eckstein et~al.}(2009)\citenamefont{Eckstein, Hackl,
  Kehrein, Kollar, Moeckel, Werner, and Wolf}}]{Eckstein+09}
\bibinfo{author}{\bibfnamefont{M.}~\bibnamefont{Eckstein}},
  \bibinfo{author}{\bibfnamefont{A.}~\bibnamefont{Hackl}},
  \bibinfo{author}{\bibfnamefont{S.}~\bibnamefont{Kehrein}},
  \bibinfo{author}{\bibfnamefont{M.}~\bibnamefont{Kollar}},
  \bibinfo{author}{\bibfnamefont{M.}~\bibnamefont{Moeckel}},
  \bibinfo{author}{\bibfnamefont{P.}~\bibnamefont{Werner}}, \bibnamefont{and}
  \bibinfo{author}{\bibfnamefont{F.}~\bibnamefont{Wolf}}, \bibinfo{journal}{The
  European Physical Journal Special Topics} \textbf{\bibinfo{volume}{180}},
  \bibinfo{pages}{217} (\bibinfo{year}{2009}).

\bibitem[{\citenamefont{Monthus}(2016)}]{Monthus16}
\bibinfo{author}{\bibfnamefont{C.}~\bibnamefont{Monthus}},
  \bibinfo{journal}{Journal of Physics A: Mathematical and Theoretical}
  \textbf{\bibinfo{volume}{49}}, \bibinfo{pages}{305002}
  (\bibinfo{year}{2016}).

\bibitem[{\citenamefont{Quito et~al.}(2016)\citenamefont{Quito, Titum, Pekker,
  and Refael}}]{Quito+16}
\bibinfo{author}{\bibfnamefont{V.~L.} \bibnamefont{Quito}},
  \bibinfo{author}{\bibfnamefont{P.}~\bibnamefont{Titum}},
  \bibinfo{author}{\bibfnamefont{D.}~\bibnamefont{Pekker}}, \bibnamefont{and}
  \bibinfo{author}{\bibfnamefont{G.}~\bibnamefont{Refael}},
  \bibinfo{journal}{Phys. Rev. B} \textbf{\bibinfo{volume}{94}},
  \bibinfo{pages}{104202} (\bibinfo{year}{2016}).

\bibitem[{\citenamefont{Pekker et~al.}(2017)\citenamefont{Pekker, Clark,
  Oganesyan, and Refael}}]{Pekker+17}
\bibinfo{author}{\bibfnamefont{D.}~\bibnamefont{Pekker}},
  \bibinfo{author}{\bibfnamefont{B.~K.} \bibnamefont{Clark}},
  \bibinfo{author}{\bibfnamefont{V.}~\bibnamefont{Oganesyan}},
  \bibnamefont{and} \bibinfo{author}{\bibfnamefont{G.}~\bibnamefont{Refael}},
  \bibinfo{journal}{Physical Review Letters} \textbf{\bibinfo{volume}{119}},
  \bibinfo{pages}{075701} (\bibinfo{year}{2017}).

\bibitem[{\citenamefont{Savitz et~al.}(2019)\citenamefont{Savitz, Peng, and
  Refael}}]{Savitz+19}
\bibinfo{author}{\bibfnamefont{S.}~\bibnamefont{Savitz}},
  \bibinfo{author}{\bibfnamefont{C.}~\bibnamefont{Peng}}, \bibnamefont{and}
  \bibinfo{author}{\bibfnamefont{G.}~\bibnamefont{Refael}},
  \bibinfo{journal}{Phys. Rev. B} \textbf{\bibinfo{volume}{100}},
  \bibinfo{pages}{094201} (\bibinfo{year}{2019}).

\bibitem[{\citenamefont{You et~al.}(2019)\citenamefont{You, Pekker, and
  Clark}}]{Yu+19}
\bibinfo{author}{\bibfnamefont{X.}~\bibnamefont{You}},
  \bibinfo{author}{\bibfnamefont{D.}~\bibnamefont{Pekker}}, \bibnamefont{and}
  \bibinfo{author}{\bibfnamefont{B.~K.} \bibnamefont{Clark}}
  (\bibinfo{year}{2019}), \bibinfo{note}{arXiv:1909.11097}.

\bibitem[{\citenamefont{Kelly et~al.}(2020)\citenamefont{Kelly, Nandkishore,
  and Marino}}]{Kelly+20}
\bibinfo{author}{\bibfnamefont{S.~P.} \bibnamefont{Kelly}},
  \bibinfo{author}{\bibfnamefont{R.}~\bibnamefont{Nandkishore}},
  \bibnamefont{and} \bibinfo{author}{\bibfnamefont{J.}~\bibnamefont{Marino}},
  \bibinfo{journal}{Nuclear Physics B} \textbf{\bibinfo{volume}{951}},
  \bibinfo{pages}{114886} (\bibinfo{year}{2020}).

\bibitem[{\citenamefont{Thomson et~al.}(2020)\citenamefont{Thomson, Magano, and
  Schir{\'o}}}]{thomson2020flow}
\bibinfo{author}{\bibfnamefont{S.~J.} \bibnamefont{Thomson}},
  \bibinfo{author}{\bibfnamefont{D.}~\bibnamefont{Magano}}, \bibnamefont{and}
  \bibinfo{author}{\bibfnamefont{M.}~\bibnamefont{Schir{\'o}}},
  \emph{\bibinfo{title}{Flow equations for disordered floquet systems}}
  (\bibinfo{year}{2020}), \eprint{2009.03186}.

\bibitem[{\citenamefont{Savitz and Refael}(2017)}]{Savitz+17}
\bibinfo{author}{\bibfnamefont{S.}~\bibnamefont{Savitz}} \bibnamefont{and}
  \bibinfo{author}{\bibfnamefont{G.}~\bibnamefont{Refael}},
  \bibinfo{journal}{Physical Review B} \textbf{\bibinfo{volume}{96}},
  \bibinfo{pages}{115129} (\bibinfo{year}{2017}).

\bibitem[{\citenamefont{Wegner}(2006)}]{Wegner06}
\bibinfo{author}{\bibfnamefont{F.}~\bibnamefont{Wegner}},
  \bibinfo{journal}{Journal of Physics A: Mathematical and General}
  \textbf{\bibinfo{volume}{39}}, \bibinfo{pages}{8221} (\bibinfo{year}{2006}).

\bibitem[{\citenamefont{Weinberg and Bukov}(2017)}]{Weinberg+17}
\bibinfo{author}{\bibfnamefont{P.}~\bibnamefont{Weinberg}} \bibnamefont{and}
  \bibinfo{author}{\bibfnamefont{M.}~\bibnamefont{Bukov}},
  \bibinfo{journal}{SciPost Phys.} \textbf{\bibinfo{volume}{2}},
  \bibinfo{pages}{003} (\bibinfo{year}{2017}).

\bibitem[{\citenamefont{Weinberg and Bukov}(2019)}]{Weinberg+19}
\bibinfo{author}{\bibfnamefont{P.}~\bibnamefont{Weinberg}} \bibnamefont{and}
  \bibinfo{author}{\bibfnamefont{M.}~\bibnamefont{Bukov}},
  \bibinfo{journal}{SciPost Phys.} \textbf{\bibinfo{volume}{7}},
  \bibinfo{pages}{20} (\bibinfo{year}{2019}).

\bibitem[{\citenamefont{Rademaker and Ortu\~no}(2016)}]{Rademaker+16}
\bibinfo{author}{\bibfnamefont{L.}~\bibnamefont{Rademaker}} \bibnamefont{and}
  \bibinfo{author}{\bibfnamefont{M.}~\bibnamefont{Ortu\~no}},
  \bibinfo{journal}{Phys. Rev. Lett.} \textbf{\bibinfo{volume}{116}},
  \bibinfo{pages}{010404} (\bibinfo{year}{2016}).

\bibitem[{\citenamefont{Rademaker et~al.}(2017)\citenamefont{Rademaker,
  Ortu\~no, and Somoza}}]{Rademaker+17}
\bibinfo{author}{\bibfnamefont{L.}~\bibnamefont{Rademaker}},
  \bibinfo{author}{\bibfnamefont{M.}~\bibnamefont{Ortu\~no}}, \bibnamefont{and}
  \bibinfo{author}{\bibfnamefont{A.~M.} \bibnamefont{Somoza}},
  \bibinfo{journal}{Annalen der Physik} pp. \bibinfo{pages}{1600322--n/a}
  (\bibinfo{year}{2017}), ISSN \bibinfo{issn}{1521-3889},
  \bibinfo{note}{1600322}.

\bibitem[{\citenamefont{Luitz et~al.}(2016)\citenamefont{Luitz, Laflorencie,
  and Alet}}]{LuitzEtALPRB16}
\bibinfo{author}{\bibfnamefont{D.~J.} \bibnamefont{Luitz}},
  \bibinfo{author}{\bibfnamefont{N.}~\bibnamefont{Laflorencie}},
  \bibnamefont{and} \bibinfo{author}{\bibfnamefont{F.}~\bibnamefont{Alet}},
  \bibinfo{journal}{Phys. Rev. B} \textbf{\bibinfo{volume}{93}},
  \bibinfo{pages}{060201(R)} (\bibinfo{year}{2016}).

\bibitem[{\citenamefont{Biroli and Tarzia}(2017)}]{BiroliTarziaPRB17}
\bibinfo{author}{\bibfnamefont{G.}~\bibnamefont{Biroli}} \bibnamefont{and}
  \bibinfo{author}{\bibfnamefont{M.}~\bibnamefont{Tarzia}},
  \bibinfo{journal}{Phys. Rev. B} \textbf{\bibinfo{volume}{96}},
  \bibinfo{pages}{201114(R)} (\bibinfo{year}{2017}).

\bibitem[{\citenamefont{Gopalakrishnan and Huse}(2019)}]{Gopalakrishnan+19}
\bibinfo{author}{\bibfnamefont{S.}~\bibnamefont{Gopalakrishnan}}
  \bibnamefont{and} \bibinfo{author}{\bibfnamefont{D.~A.} \bibnamefont{Huse}},
  \bibinfo{journal}{Physical Review B} \textbf{\bibinfo{volume}{99}},
  \bibinfo{pages}{134305} (\bibinfo{year}{2019}).

\bibitem[{\citenamefont{Garc{\'\i}a-Garc{\'\i}a and Tezuka}(2019)}]{Garcia+19}
\bibinfo{author}{\bibfnamefont{A.~M.} \bibnamefont{Garc{\'\i}a-Garc{\'\i}a}}
  \bibnamefont{and} \bibinfo{author}{\bibfnamefont{M.}~\bibnamefont{Tezuka}},
  \bibinfo{journal}{Physical Review B} \textbf{\bibinfo{volume}{99}},
  \bibinfo{pages}{054202} (\bibinfo{year}{2019}).

\bibitem[{\citenamefont{Wurtz et~al.}(2020)\citenamefont{Wurtz, Claeys, and
  Polkovnikov}}]{Wurtz+20}
\bibinfo{author}{\bibfnamefont{J.}~\bibnamefont{Wurtz}},
  \bibinfo{author}{\bibfnamefont{P.~W.} \bibnamefont{Claeys}},
  \bibnamefont{and}
  \bibinfo{author}{\bibfnamefont{A.}~\bibnamefont{Polkovnikov}},
  \bibinfo{journal}{Phys. Rev. B} \textbf{\bibinfo{volume}{101}},
  \bibinfo{pages}{014302} (\bibinfo{year}{2020}).

\bibitem[{\citenamefont{{Crowley} and
  {Chandran}}(2019)}]{CrowleyChandranArxiv19}
\bibinfo{author}{\bibfnamefont{P.~J.~D.} \bibnamefont{{Crowley}}}
  \bibnamefont{and}
  \bibinfo{author}{\bibfnamefont{A.}~\bibnamefont{{Chandran}}},
  \bibinfo{journal}{arXiv e-prints} \bibinfo{eid}{arXiv:1910.10812}
  (\bibinfo{year}{2019}), \eprint{1910.10812}.

\bibitem[{\citenamefont{Rosso et~al.}(2020)\citenamefont{Rosso, Iemini,
  Schir{\`o}, and Mazza}}]{rosso2020dissipative}
\bibinfo{author}{\bibfnamefont{L.}~\bibnamefont{Rosso}},
  \bibinfo{author}{\bibfnamefont{F.}~\bibnamefont{Iemini}},
  \bibinfo{author}{\bibfnamefont{M.}~\bibnamefont{Schir{\`o}}},
  \bibnamefont{and} \bibinfo{author}{\bibfnamefont{L.}~\bibnamefont{Mazza}},
  \bibinfo{journal}{arXiv preprint arXiv:2007.12044}  (\bibinfo{year}{2020}).

\bibitem[{\citenamefont{Prasad and Garg}(2020)}]{prasad2020enhanced}
\bibinfo{author}{\bibfnamefont{Y.}~\bibnamefont{Prasad}} \bibnamefont{and}
  \bibinfo{author}{\bibfnamefont{A.}~\bibnamefont{Garg}},
  \bibinfo{journal}{arXiv preprint arXiv:2010.12485}  (\bibinfo{year}{2020}).

\end{thebibliography}

\end{document}